\documentclass[dvipdfmx,manuscript]{aastex61}
\usepackage{graphicx}
\usepackage{bm}

\newcommand\aastex{AAS\TeX}

\received{July 1, 2016}
\revised{September 27, 2016}
\accepted{\today}
\submitjournal{ApJ}

\shorttitle{\aastex\ Semi-Analytical Model for Wind-Fed Black Hole High-Mass X-ray Binaries}
\shortauthors{Yaji, Yamada, \& Masai}

\begin{document}

\title{Semi-Analytical Model for Wind-Fed Black Hole High-Mass X-ray Binaries \\ -- State Transition Triggered by Magnetic Fields from the Companion Star --}
\author{Kentaro Yaji}

\author{Shinya Yamada}

\author{Kuniaki Masai}
\affiliation{Department of Physics, Tokyo Metropolitan University, Minami-Osawa 1-1, Hachioji, Tokyo 192-0397, Japan}

\begin{abstract}
We propose a mechanism of state transition in wind-fed black hole binaries (high-mass X-ray binaries) such as Cyg X-1 and LMC X-1.
Modeling a line-driven stellar wind from the companion by two-dimensional hydrodynamical calculations, we investigate the processes of wind capture by and accretion onto the black hole. We assume that the wind acceleration is terminated at the He\,II ionization front because ions responsible for line-driven acceleration are ionized within the front, i.e. He III region. It is found that the mass accretion rate inferred from the luminosity is remarkably smaller than the capture rate. Considering the difference, we construct a model for the state transition based on the accretion flow being controlled by magneto-rotational instability. The outer flow is torus like, and plays an important role to trigger the transition. The model can explain why state transition does occur in Cyg X-1, while not in LMC X-1. Cyg X-1 exhibits a relatively low luminosity, and then the He\,II ionization front is located and can move between the companion and black hole, depending on its ionizing photon flux. On the other hand, LMC X-1 exhibits too high luminosity for the front to move considerably; the front is too close to the companion atmosphere. The model also predicts that each state of high-soft or low-hard would last fairly long because the luminosity depends weakly on the wind velocity. In the context of the model, the state transition is triggered by a fluctuation of the magnetic field when its amplitude becomes comparable to the field strength in the torus-like outer flow.

\end{abstract}

\section{Introduction} \label{sec:intro}

It is well known that black hole (BH) X-ray binaries have different X-ray spectral states (Done et al. 2007). 
The main two states are high-soft (HS) state  and low-hard (LH) state. 
In the HS state, X-ray emission is dominated by low-energy photons of multi-color blackbody radiation which comes from a geometrically thin disk (Shakura \& Sunyaev 1973; Done 2010). 
In the LH state, X-ray emission is dominated by high-energy photons with a power-law spectrum which originates likely from a hot corona or accretion flow (Yuan \& Narayan 2014).

The relations between the spectral states and the accretion flows are well studied in low-mass X-ray binaries (LMXBs) which are ``disk-fed'' objects.
 In these systems, the gas of the companion envelope overflows through the Lagrange point toward the BH.
  Therefore, the radial velocity is so low ($\sim 10^6$ cm s$^{-1}$) that the accretion disk can be formed on the scale of $\sim 10^{11}$ cm and the density is so high that the outer edge of the disk can become optically thick. 
On the other hand, in high-mass X-ray binaries (HMXBs) which are ``wind-fed'' objects, the OB-type star companion blows fast ($\sim 10^8$ cm s$^{-1}$) stellar wind with a mass loss rate $\sim 10^{-6} M_{\odot}$ yr$^{-1}$.
 Since the physical properties of the accreting gas are different, it is not obvious that the accretion physics in HMXBs is same as LMXBs. 

 There are some observational facts on the difference between the two systems. 
 Cyg X-1, which belongs to HMXBs, shows both HS state and LH state. 
 It stays in the LH state for a long time. 
 The bolometric luminosity between the two states changes only by a factor of $\sim$3-4 (Nowak et al. 2012; Zdziarski et al. 2002). 
 Another famous example of a HMXB is LMC X-1. 
 It has stayed in the HS state since its discovery. The luminosity varies by a factor of $< 4$ (Wilms et al. 2001; Ruhlen et al. 2011). 
 On the other hand, in LMXBs such as GX~339-4, the luminosity often changes by more than two orders of magnitude during an outburst (Belloni et al. 2005; Nowak et al. 2012). 
Since OB-type stars continuously blow the stellar wind, the persistent property of HMXBs can be explained by the wind-fed accretion.

In HMXBs, the strong X-ray radiation from the black hole could change the ionization state of the wind material via photoionization. 
The effect has been well studied in theory (cf., Hatchett \& McCray 1977; Masai 1984; MacGregor \& Vitello 1982; \u{C}echura \& Hadrava 2015). 
They are confirmed by observations; e.g., the orbital variability of absorption and the modulation of a P-Cygni profile (Grinberg et al. 2015; Gies et al. 2008). 
On the other hand, the X-ray spectra in the HS and LH states have been extensively studied in both theory and observation. 
The gross properties of the accretion disk in the two states are understood as a geometrically thin disk and hot corona (e.g., Yamada et al. 2013ab). 
However, the mechanism of state transitions is still unclear. 
The process of how the wind materials eventually turn into a part of the accretion flow and affect the spectral states need to be studied. 
 
In this paper, we focus on the state transition mechanism in BH HMXBs.
We examine the wind capture process by conducting 2-D hydrodynamical calculations (\software{PLUTO}; Mignone et al. 2007). 
It approximately includes X-ray photoionization effects.
Then, we analytically connect the captured wind with the outer accretion flow.
We assume that magneto-rotational instability (MRI) operates to transport the angular momentum of the flow. 
The criterion for MRI on-set is suggested by previous works (Pessah \& Psaltis 2005; Begelman \& Pringle 2007; Begelman et al. 2015).
It plays a key role in elucidating a link between the net accretion rate and the state transitions. 

In Section 2, we investigate the wind-fed process and construct an analytical model for estimations of the physical properties of the outer accretion flow. 
In Section 3, based on our model, we discuss the state transition mechanism which can explain some observational features of Cyg X-1 and LMC X-1. 
They are summarized in Section 4.
 
 \section{model and calculations}\label{sec:2}
 We examine the wind capture process and construct a model connecting the wind with the accretion flow.
In our calculations, the wind velocity outside the He$\rm{{I\hspace{-.1em}I}}$ ionization front follows an empirical profile reproduced by the CAK model (Castor et al. 1975), 
while the wind acceleration inside the front is terminated by the X-ray photoionization (Masai 1984).
 We expect that the captured wind gas forms a Keplerian rotating torus-like flow because of the fast sound speed of shocked matters and asymmetrical effects such as the orbital motion.
We also expect that a viscous process must operate for the extraction of angular momentum from the flow.
In our model, magneto-rotational instability (MRI)  leads to the accretion of the torus-like flow, and determines the accretion rate.
 During this accretion process, the density increases, and then the flow would become optically thick and enter the multi-color disk regime at some radius.

\subsection{Wind-fed processes}\label{sec:2.1}
 Using the \software{PLUTO} hydrodynamical calculation code (Mignone et al. 2007), we examine an axisymmetric wind capture process.
 BH gravity and centrifugal force due to the rotation of the companion star enhance the stellar wind along the line connecting the BH and the companion (Friend \& Castor 1982; Shimizu et al. 2012).
Therefore, we consider that the wind does not diverge toward the vertical direction (i.e. cylindrical wind).
Our calculations are conducted on a two-dimensional cylindrical coordinate ($R,\varphi$) grid and we set the BH as the origin.
 For all calculations, we adopt a grid of 512$\times$512 cells. 
 The basic equations are 
 \begin{equation}
 \frac{\partial \rho}{\partial t}+\nabla \cdotp(\rho\bm{v})=0,
\label{eq:3}
\end{equation}
 \begin{equation}
\frac{\partial \rho \bm{v}}{\partial t}+\nabla \cdotp(\rho \bm{v}\bm{v}+P)+\rho \nabla \Psi=0,
  \label{eq:4}
\end{equation}
 \begin{equation}
 \frac{\partial E}{\partial t}+\nabla \cdotp [(E+P)\bm{v}]+\rho \bm{v}\cdotp\nabla \Psi=0,
 \label{eq:5}
\end{equation}
 \begin{equation}
 E=P/(\gamma-1)+\frac{1}{2}\rho v^2,
  \label{eq:6}
\end{equation}
 \begin{equation}
\Psi=\Psi_{BH}-\Psi_{\rm{accel}}, 
\label{eq:8}
\end{equation}
where $\rho, \bm{v}, P, E$ and $\gamma = 5/3$ are density, velocity, pressure, energy, and adiabatic index, respectively. 
We adopt an ideal gas equation of state.
We use a Newtonian gravitational potential of BH $\Psi_{BH} $ and wind acceleration potential
\begin{equation}
\Psi_{\rm{accel}} (r) =\frac{\psi}{r}\;\; (R>\mbox{He$\rm{{I\hspace{-.1em}I}}$ ionization front}),\\
\Psi_{\rm{accel}} (r) =0\;\;\; (R<\mbox{He$\rm{{I\hspace{-.1em}I}}$ ionization front}),
\label{eq:c}
\end{equation}
where $r$ is the distance from the companion star.
$\psi$ is a constant, which approximately reproduces the empirical velocity profile of line-driven wind 
\begin{equation}
V_{\rm{empirical}}(r) =V_{\infty}\sqrt{1-\frac{r_*}{r}}, 
\label{eq:9}
\end{equation}
where $V_{\infty}$ is the terminal wind velocity of an isolated OB-type star.
We suppose that the radiation from the companion star is artificially effective on the wind acceleration as Eq. (\ref{eq:c}).

The effect of X-ray irradiation on the line-driven wind in HMXBs has been investigated by some studies.
MacGregor \& Vitello (1982) and Watanabe et al. (2006) demonstrated that the wind velocity near the X-ray source becomes lower than that expected by the CAK model.
Masai (1984) showed that ions responsible for the acceleration are protected by a small fraction of He$\rm{{I\hspace{-.1em}I}}$ left near the boundary of the almost totally ionized helium region.
Here, note that hydrogen is the most abundant in the wind, but it is completely ionized. 

 In this paper, we consider that there are two kinds of He$\rm{{I\hspace{-.1em}I}}$ ionization front correlated with X-ray spectral states.
  In the HS state, the low energy photons dominate in the spectrum.
  Therefore, we adopt the ionization front as the boundary where the optical depth becomes of order unity for photons of the He$\rm{{I\hspace{-.1em}I}}$ ionization energy $\sim 54.4$ eV.
 In the LH state, the ionization front can spread over a larger scale.
    This is because most photons have much higher energy than the He$\rm{{I\hspace{-.1em}I}}$ ionization energy and the cross section of photoelectric absorption decreases with energy as $\propto E^{-3}$. 
 Following Masai (1984), we assume the X-ray spectrum as 
 \begin{equation}
\left(dL_X/dE \right)dE\propto E^{-\alpha_X}\exp \left(E/E_f\right)EdE,
 \label{eq_v3:1}
 \end{equation}
where $L_X$ and $E_f$ are X-ray luminosity and folding energy.
We take $\alpha_X =1$ for both states and this form approximately corresponds with the spectrum of bremsstrahlung.
We assume that Eq. (\ref{eq_v3:1}) holds all over the range including the  He$\rm{{I\hspace{-.1em}I}}$ ionization energy.
 For the HS state, the multi-color blackbody shows $\alpha_X =2/3$, while for the LH state, the observed X-ray spectrum shows $\alpha_X\sim1$.
  Since both states represent $\alpha\approx 1$, we believe that our assumption is reasonable.
 
 Adopting $\alpha_X=1$, we can write the He$\rm{{I\hspace{-.1em}I}}$ ionization fronts as the radii from the X-ray source so that
 \begin{equation}
R_{\rm{S}}=1.56\times10^{12} \left(\frac{n}{10^{11} \;\rm{cm^{-3}}}\right)^{-2/3} 
\\
\times \left(\frac{L_X}{10^{37}\;\rm{erg}}\right)^{1/3} \left(\frac{E_f}{2\; \rm{keV}}\right)^{-1/3} \;\;\rm{cm}, 
 \label{eq:1}
 \end{equation}
 
 \begin{equation}
R_{\rm{He}}= 5.15\times10^{11}\left(\frac{n}{10^{11} \;\rm{cm^{-3}}}\right)^{-3/5} 
\\
\times \left(\frac{L_X}{10^{37}\;\rm{erg}}\right)^{2/5} \left(\frac{E_f}{100\; \rm{keV}}\right)^{-2/5} \;\;\rm{cm}, 
\label{eq:2}
\end{equation}
 (Masai 1984) where $R_{\rm{S}}$ and $R_{\rm{He}}$ denote ionization fronts in HS state and in LH state, respectively. $n$ is number density.  
Since the radii of the front are different between the two states, the terminal velocities differ.
The X-ray luminosity and the folding energy are listed in Table \ref{tab_v4:1}.

\begin{deluxetable}{cccc}[h!]
\tablecaption{
Luminosity, folding energy, and mass accretion rate assumed in the simulation.
$\dot{M}_{\rm{expected}}$ is expected from the luminosity with the radiative efficiency $\epsilon = 0.1$.
 \label{tab_v4:1}}
\tablecolumns{6}
\tablewidth{0pt}
\tablehead{
 \colhead{Source name}&\colhead{ $L_X$ [erg/s]}&\colhead{$E_f$ [keV]}&\colhead{$\dot{M}_{\rm{expected}}$ [g s$^{-1}$]}
}
\startdata
 Cyg X-1 (HS)&$3.6\times 10^{37}$&2&4.0$\times10^{17}$ \\ 
 Cyg X-1 (LH)&$ 0.9\times10^{37}$&100&1.0$\times10^{17}$ \\ 
 LMC X-1 (HS)&2.4$\times 10^{38}$&2&2.7$\times10^{18}$\\ 
\enddata
\end{deluxetable}

In all calculations, we adopt an isotropic wind of the sound speed $c_{\rm{s}}= 4\times10^7$ cm s$^{-1}$ on the surface of companion.
It  is much higher than the typical value of OB-type stars, but the effect on the velocity distribution can be negligible.
This is because the wind velocity can become $\sim 10^8$cm s$^{-1}$ $>c_{\rm{s}}$ near the surface.
We adopt the terminal velocity $V_{\infty}\sim1.5\times 10^8$ cm s$^{-1}$ and the initial wind velocity $v_0=10^7$ cm s$^{-1}$.
Since the companion of LMC X-1 is larger than or comparable to that of Cyg X-1, we adopt the higher mass loss rate $\dot{M}_*$ for LMC X-1.
We assume the companion star gravity is negligible. 
Other parameters are listed in Table \ref{tab_v4:2}.

\startlongtable
\begin{deluxetable*}{ccccc}
\tablecaption{
Input parameters for the simulation. 
$M_X$, $r_*$, $\dot{M}_*$ and $d$ denoting black hole mass, radius of OB-type star, mass loss rate, and binary separation, respectively. 
\label{tab_v4:2}}
\tablehead{
\colhead{Source name}&\colhead{$M_X \;[M_\odot]$}&\colhead{$r_* \;$[cm]}&$\dot{M}_*$ [$M_{\odot} \rm{yr}^{-1}$]/ [g s$^{-1}$]&\colhead{d [cm]}
}
\startdata
Cyg X-1&15&$1.0\times10^{12}$&$1.25\times10^{-6}$/$8.3\times10^{19}$&$3.0\times10^{12}$\\
 LMC X-1 &  10 & $1.0\times10^{12}$&$6\times10^{-6}$/$4.0\times10^{20}$&$2.5\times10^{12}$\\ 
\enddata
\tablecomments{The black hole masses and the binary separations are chosen to be consistent with Orosz et al. (2009, 2011).}
\end{deluxetable*}



 \begin{figure*}[h]
 \begin{center}
\gridline{
 \fig{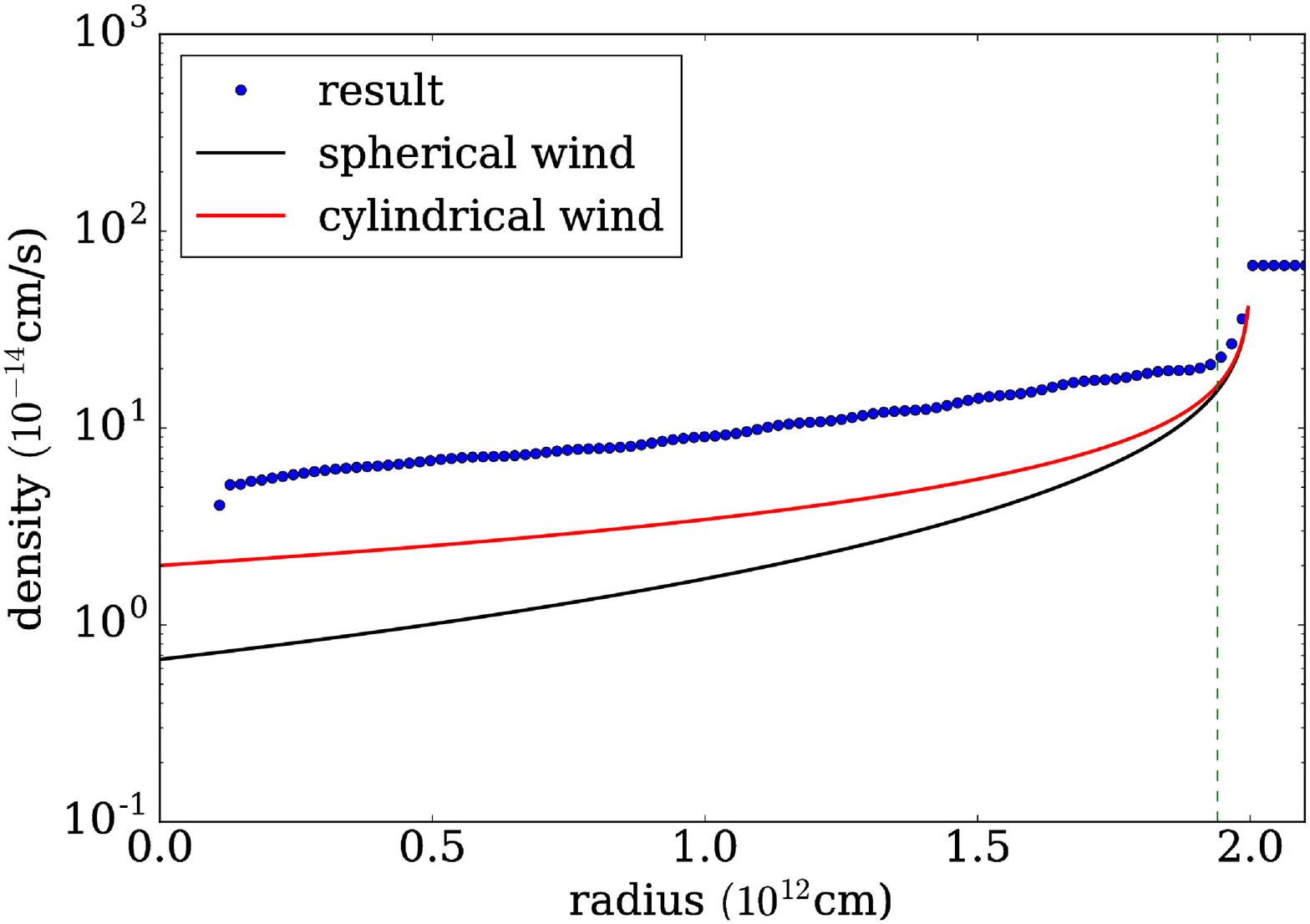}{0.5\textwidth}{(a)}
  \fig{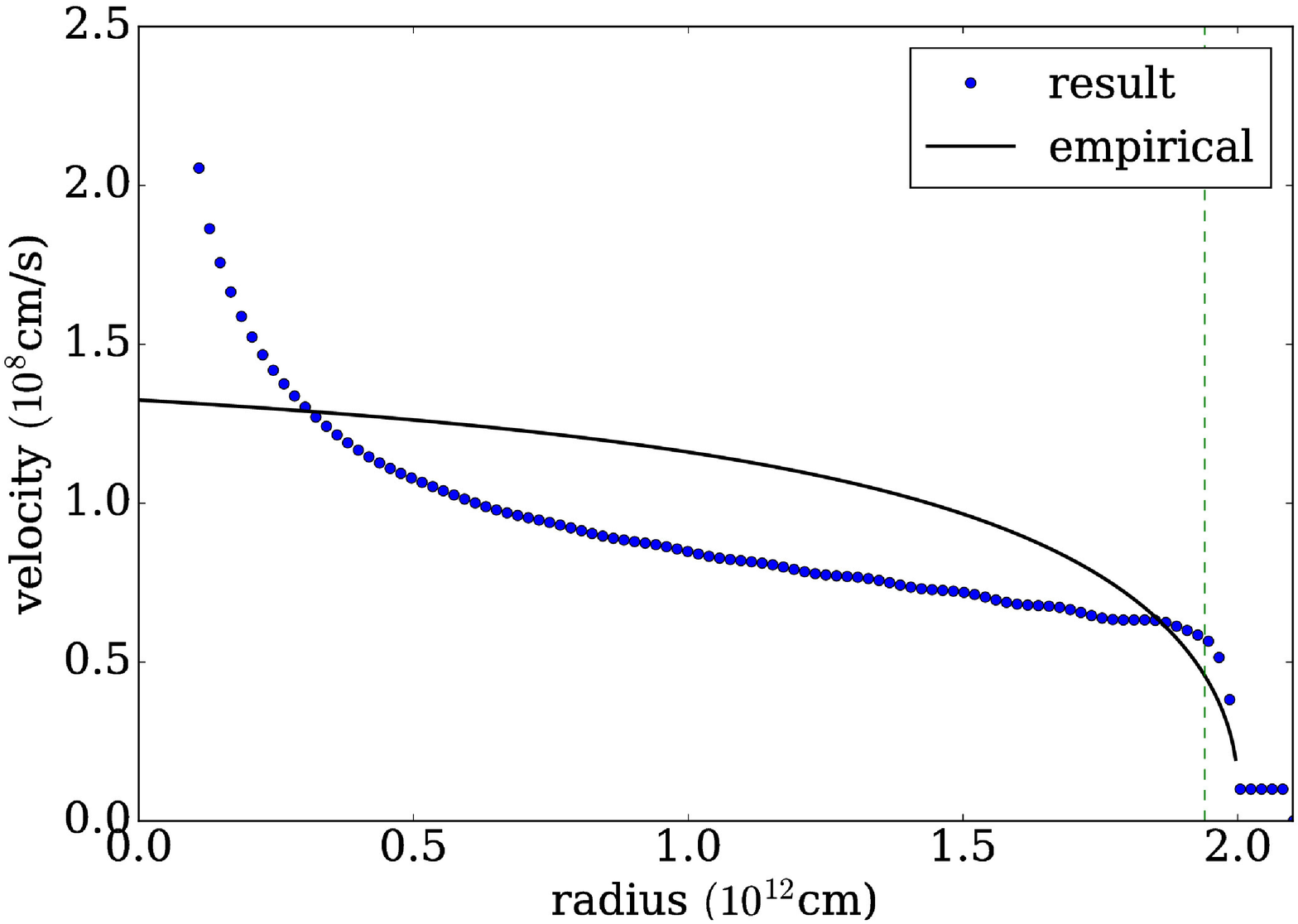}{0.5\textwidth}{(b)}
  }
\gridline{
   \fig{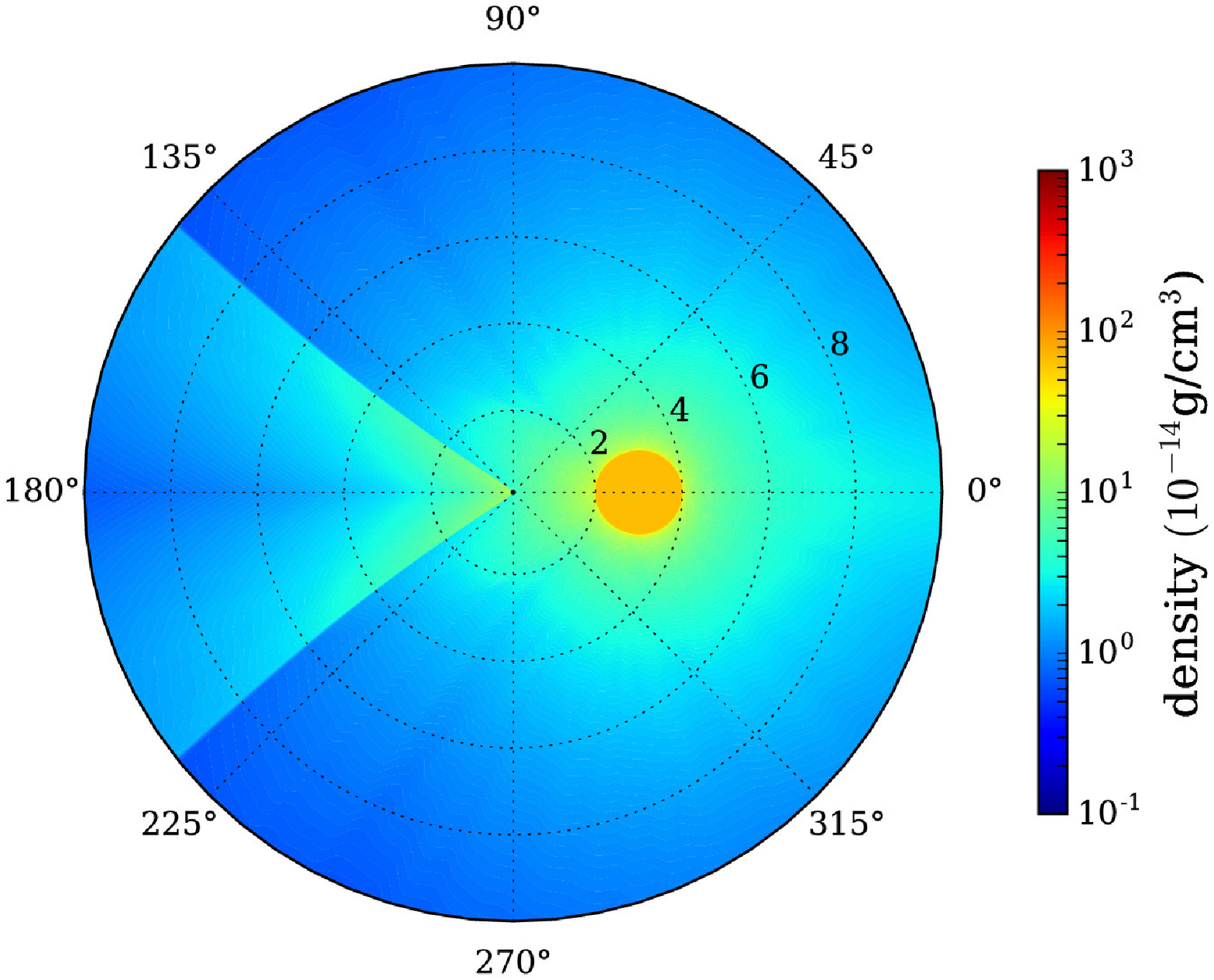}{0.5\textwidth}{(c)}
  \fig{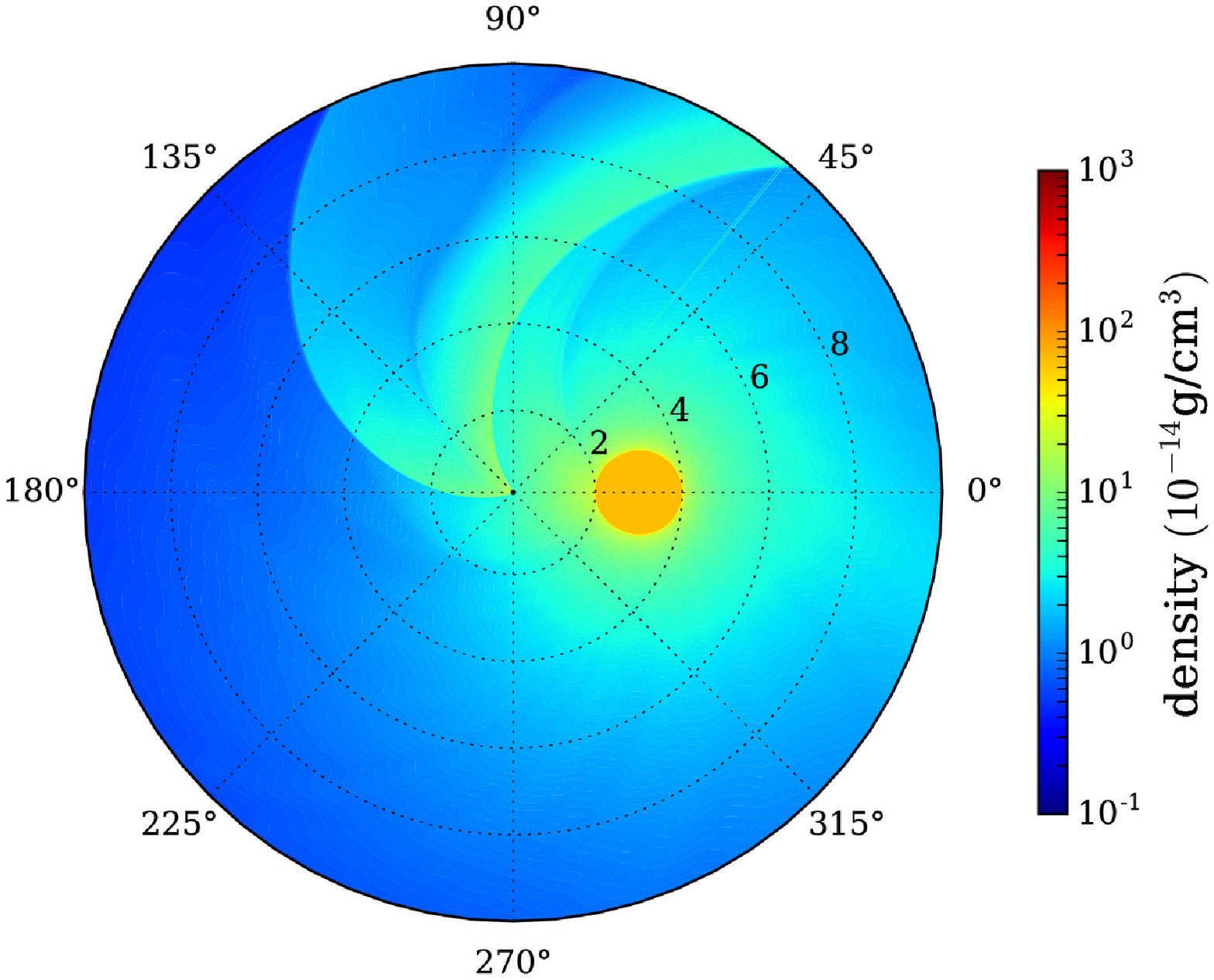}{0.5\textwidth}{(d)} 
  }  
  \caption{The results for Cyg X-1 in HS state are shown: (a) density distribution along the line connecting the BH and the companion, (b)  velocity distribution, (c) density distribution including all computational area for 1024$\times$1024 cells and (d) density distribution for a rotational frame.
In the panel (a) and (b), the positions of the BH and the surface of companion star are located at the origin and $2.0 \times 10^{12}$ cm respectively.
The green vertical dotted-line shows the position of the He$\rm{{I\hspace{-.1em}I}}$ ionization front.
In the panel (a), the blue dots, the red and black lines show the results, analytical distribution in the case of cylindrical wind and spherical wind, respectively. 
In the panel (b), the blue dots and black line show the results and empirical velocity distribution.
}
 \label{fig:1}
\end{center}
\end{figure*}
  
 \begin{figure*}[h]
 \begin{center}
 \gridline{
 \fig{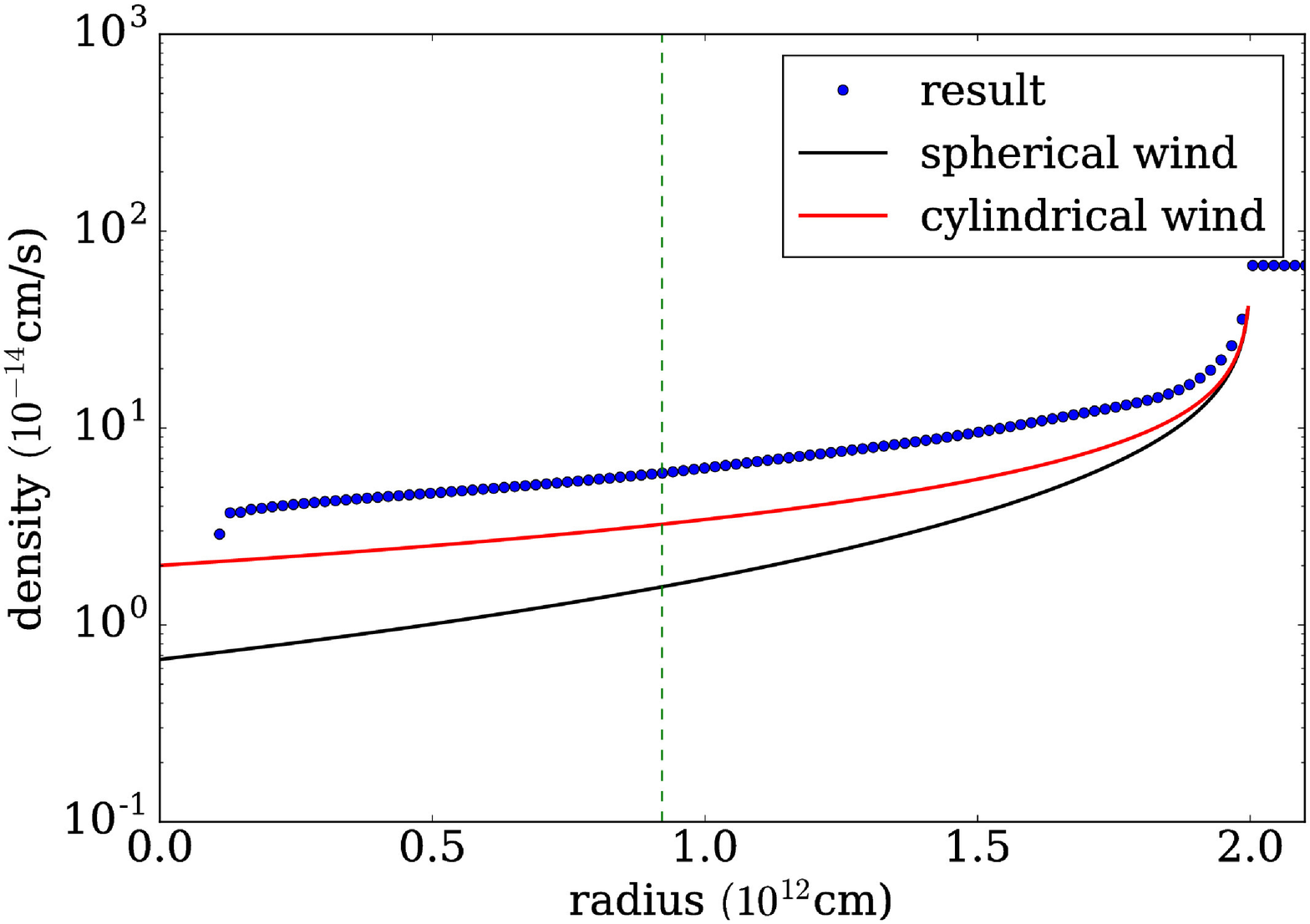}{0.5\textwidth}{(a)}
  \fig{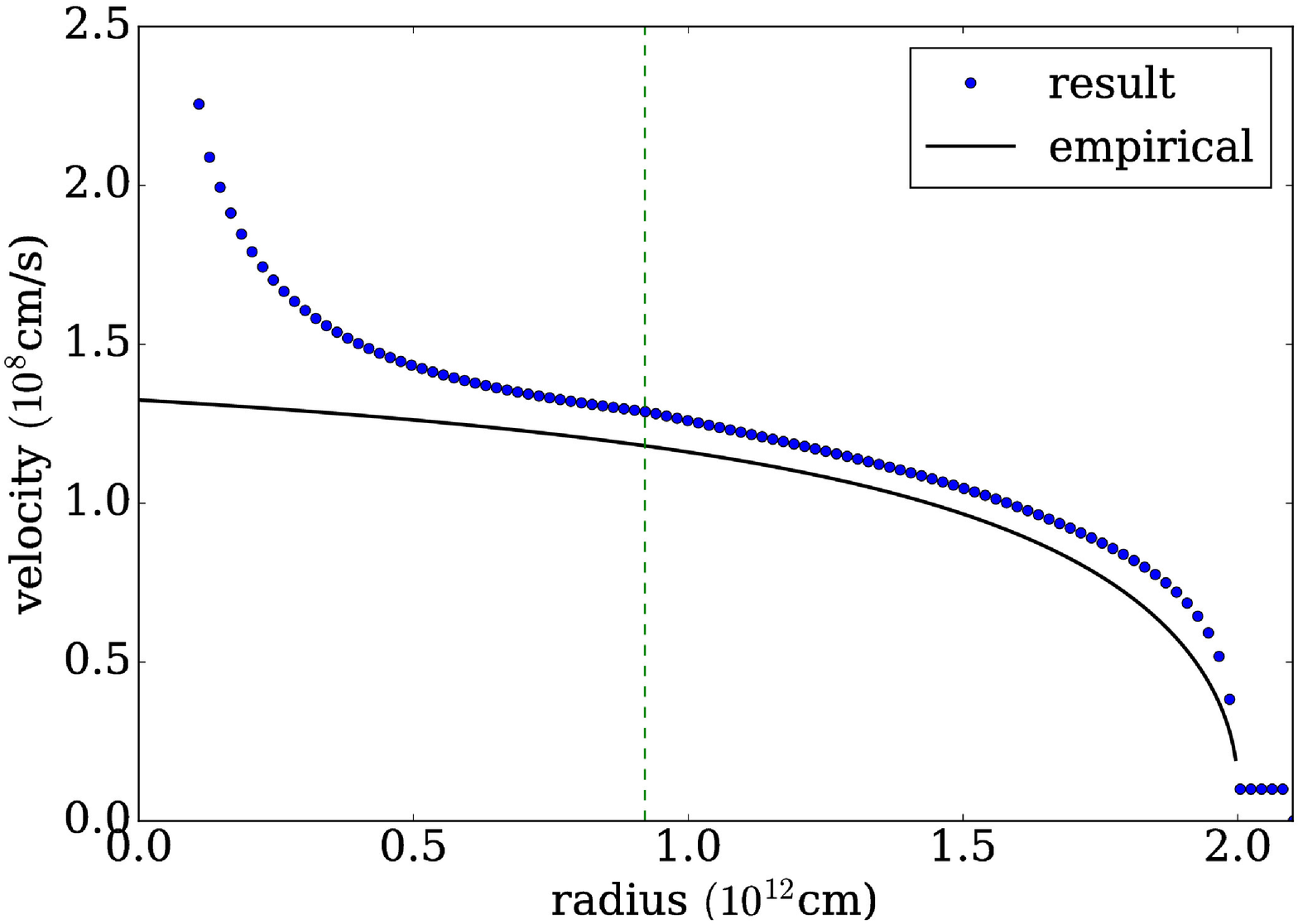}{0.5\textwidth}{(b)}
}
 \gridline{
  \fig{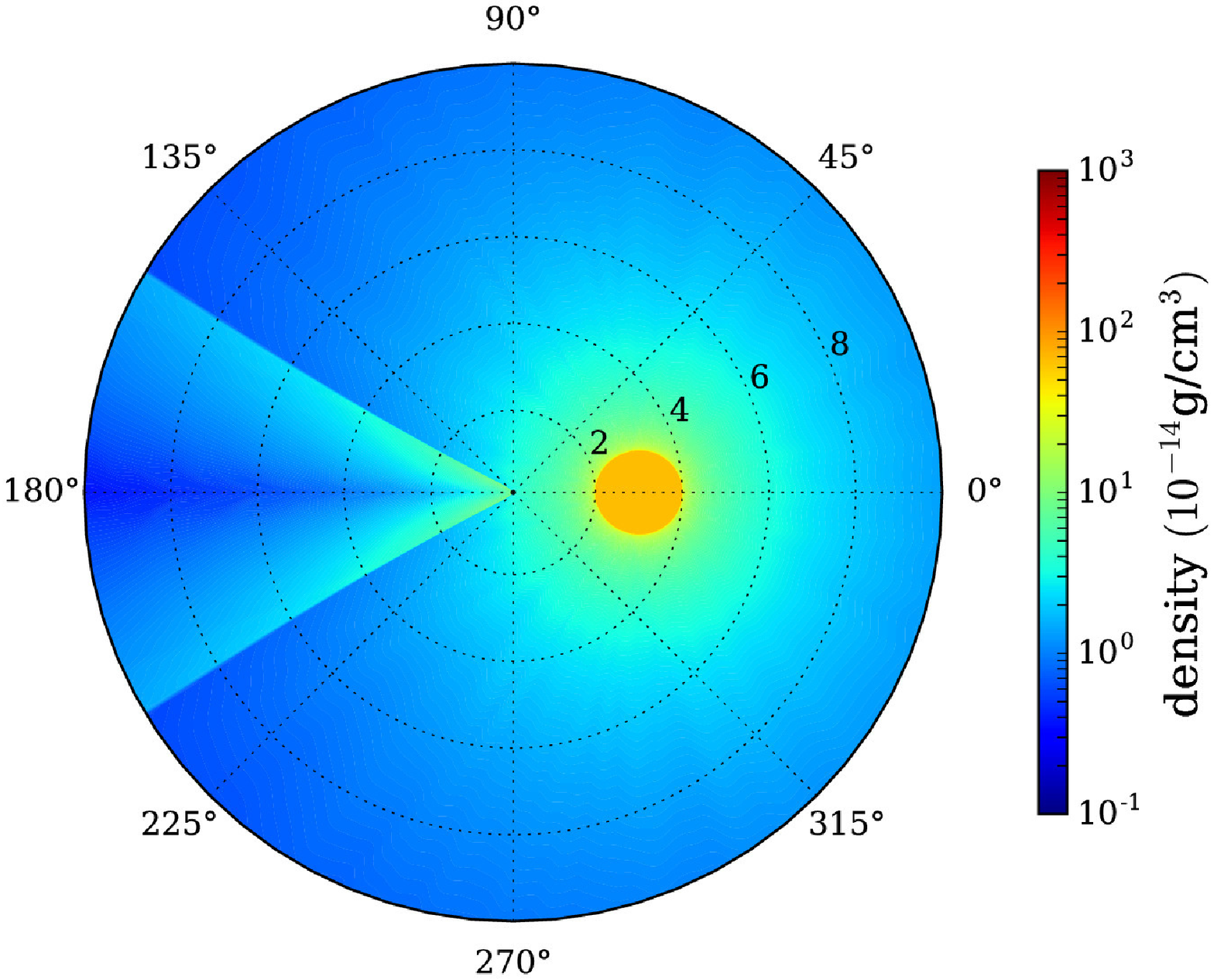}{0.5\textwidth}{(c)}
  \fig{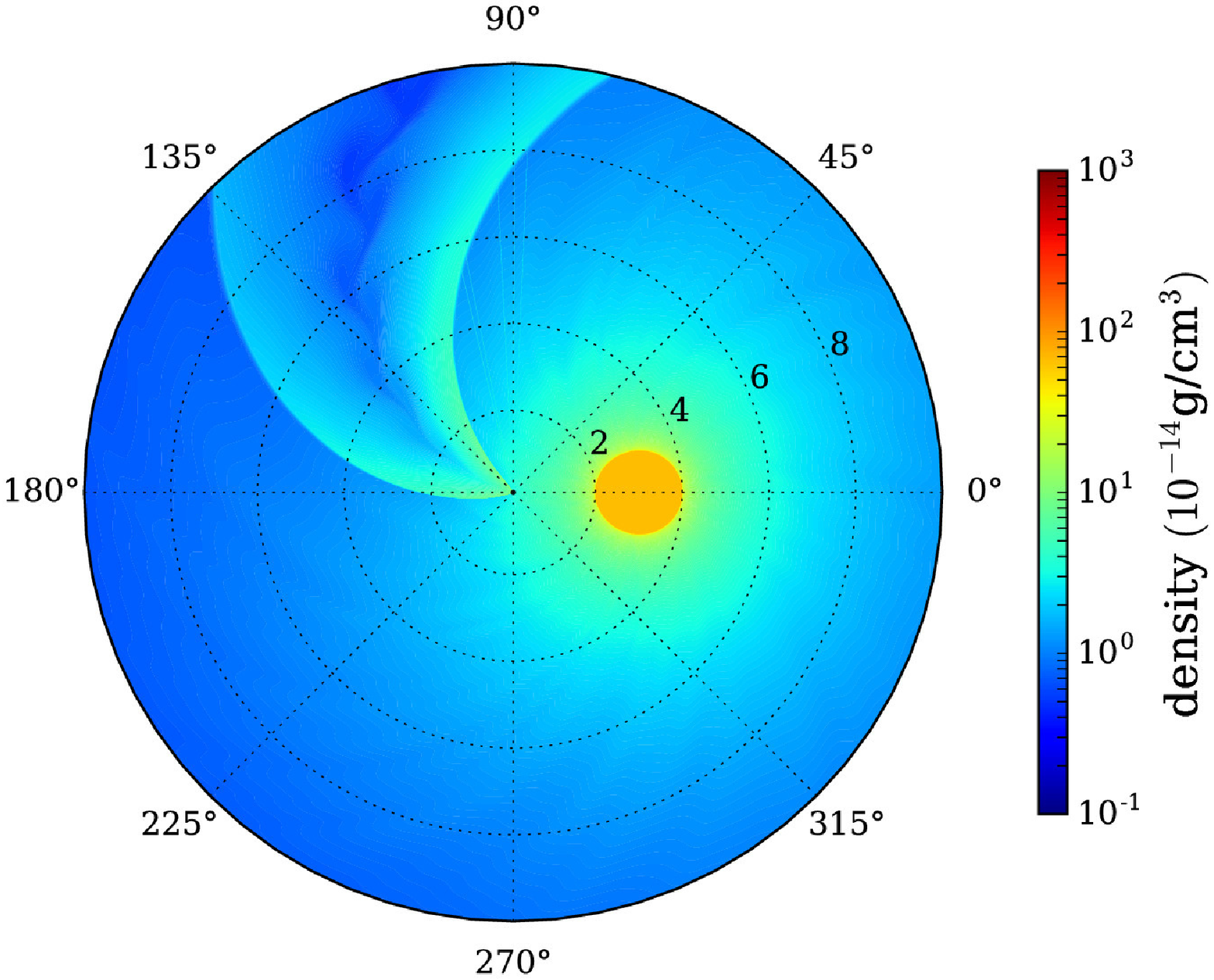}{0.5\textwidth}{(d)} 
}
  \caption{
  The same as Figure \ref{fig:1} except for LH state.
}
 \label{fig:2}
\end{center}
\end{figure*}

\begin{figure*}[h]
 \begin{center}
\gridline{
  \fig{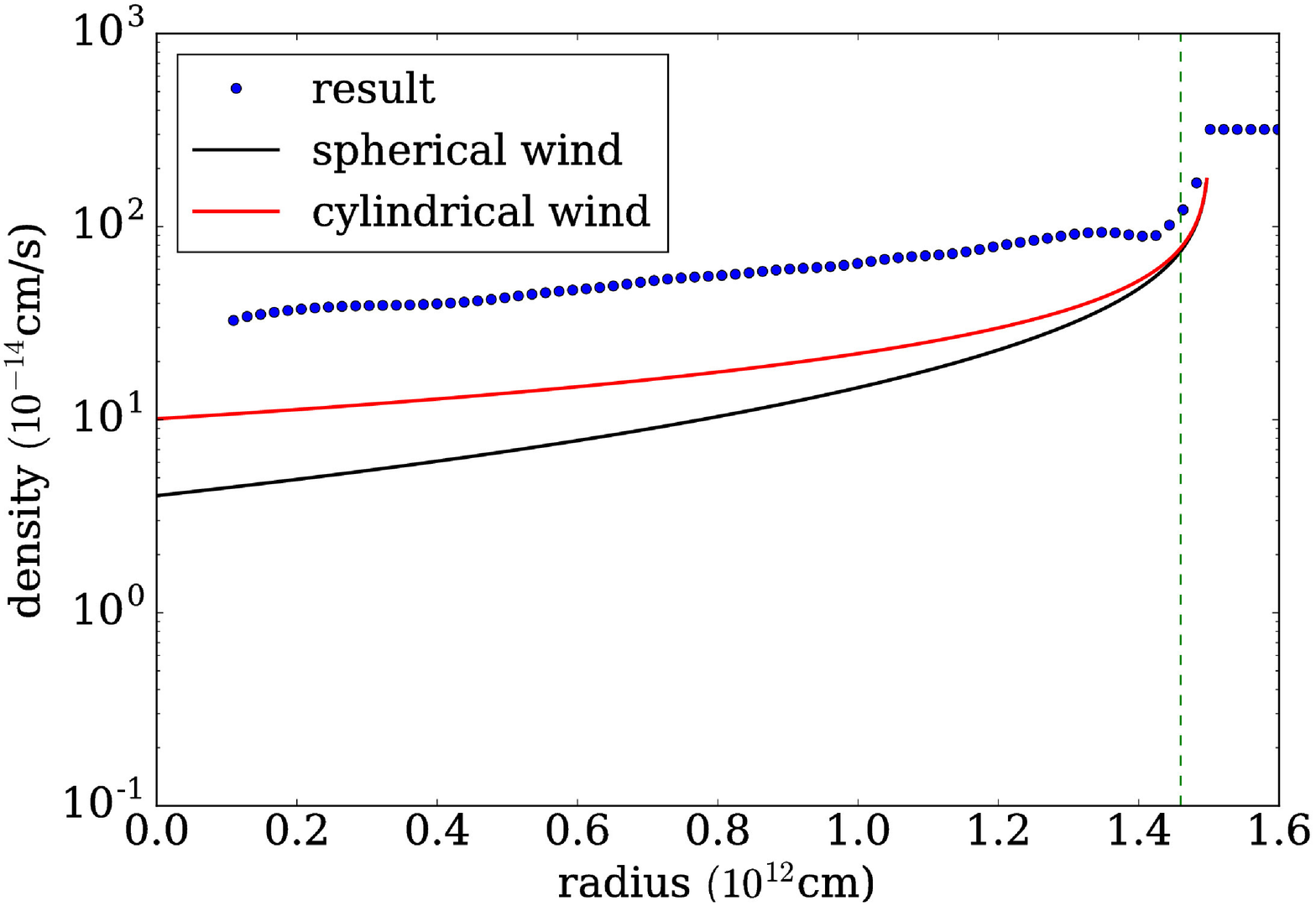}{0.5\textwidth}{(a)}
  \fig{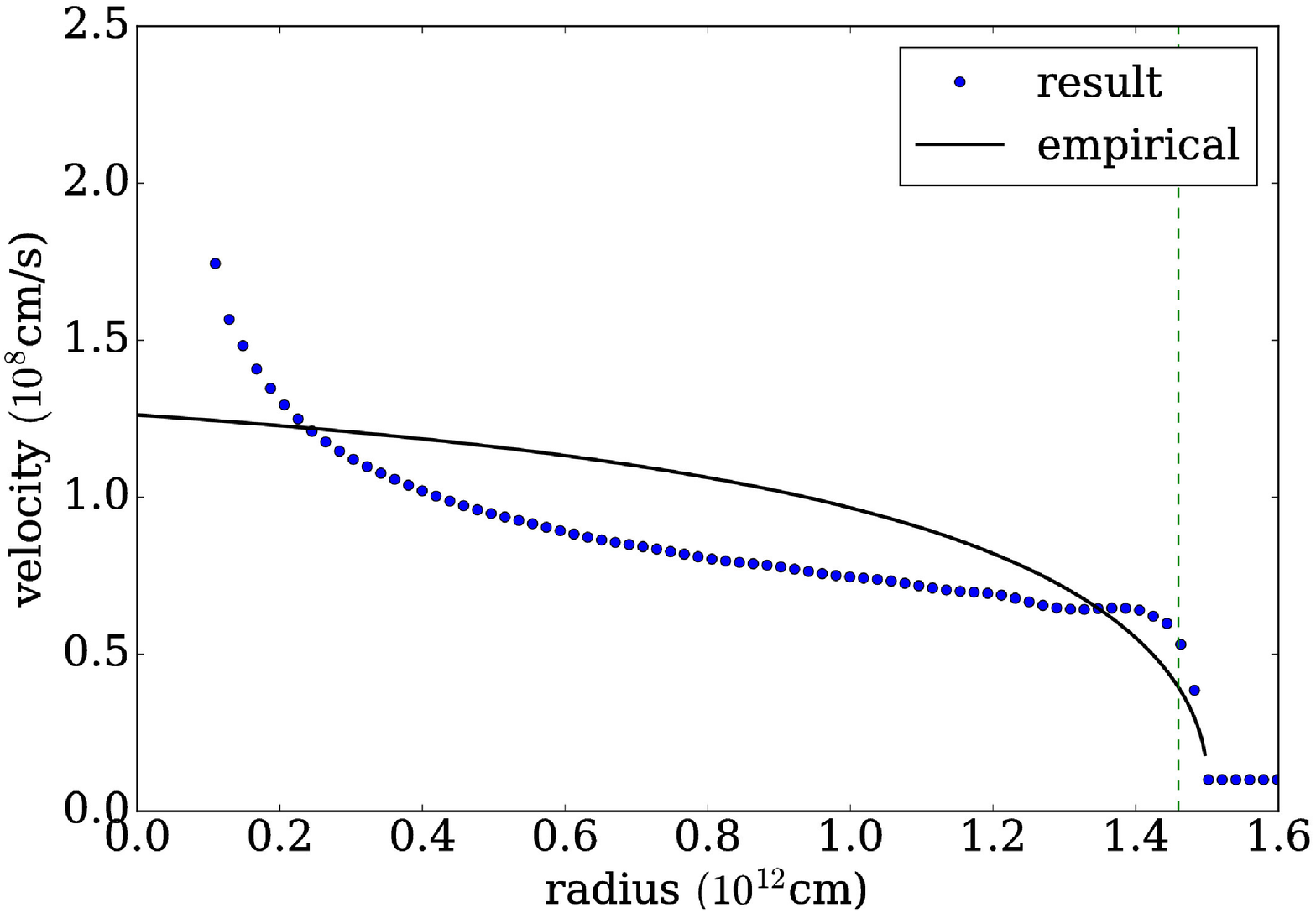}{0.5\textwidth}{(b)}
  }
\gridline{
  \fig{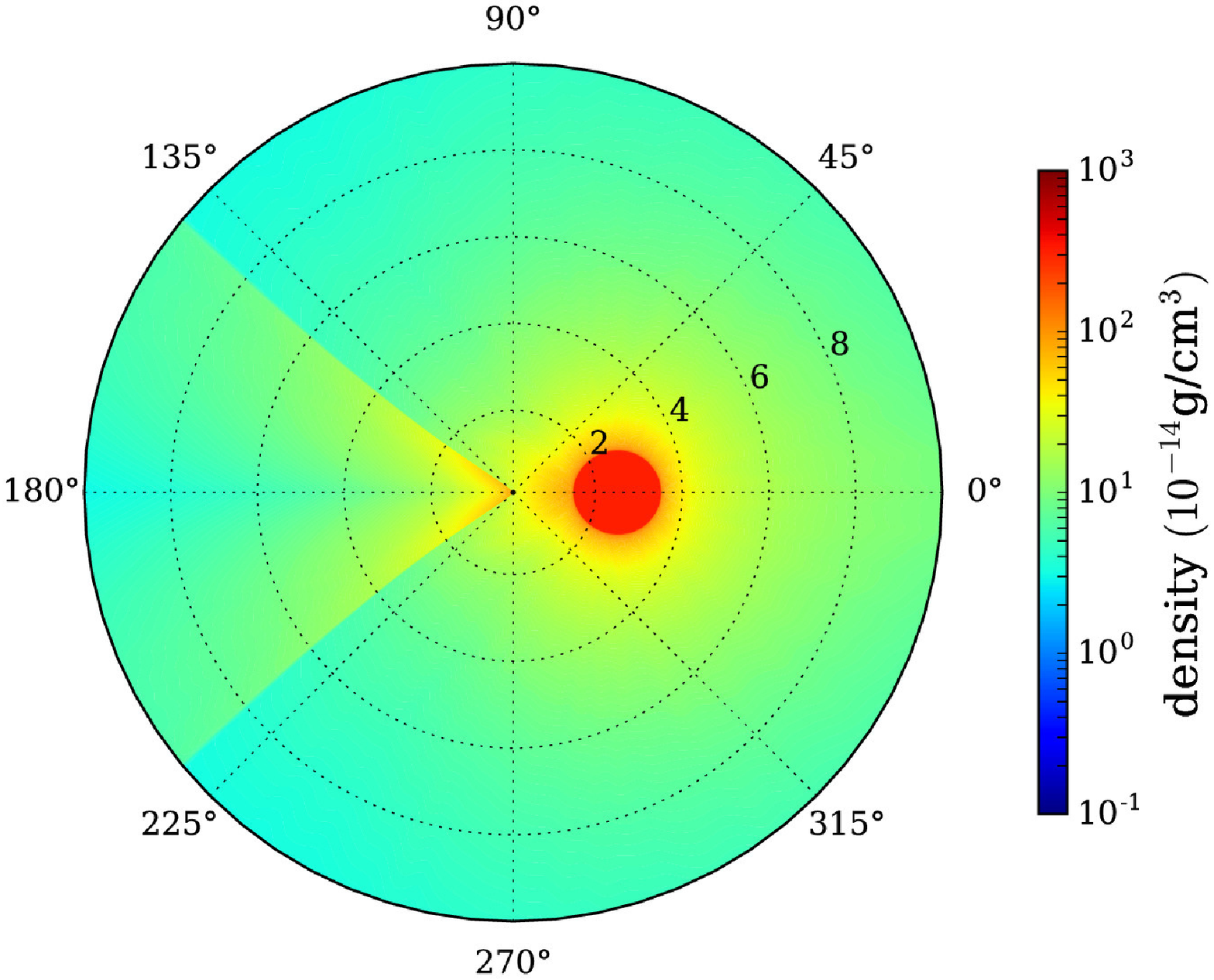}{0.5\textwidth}{(c)}
  \fig{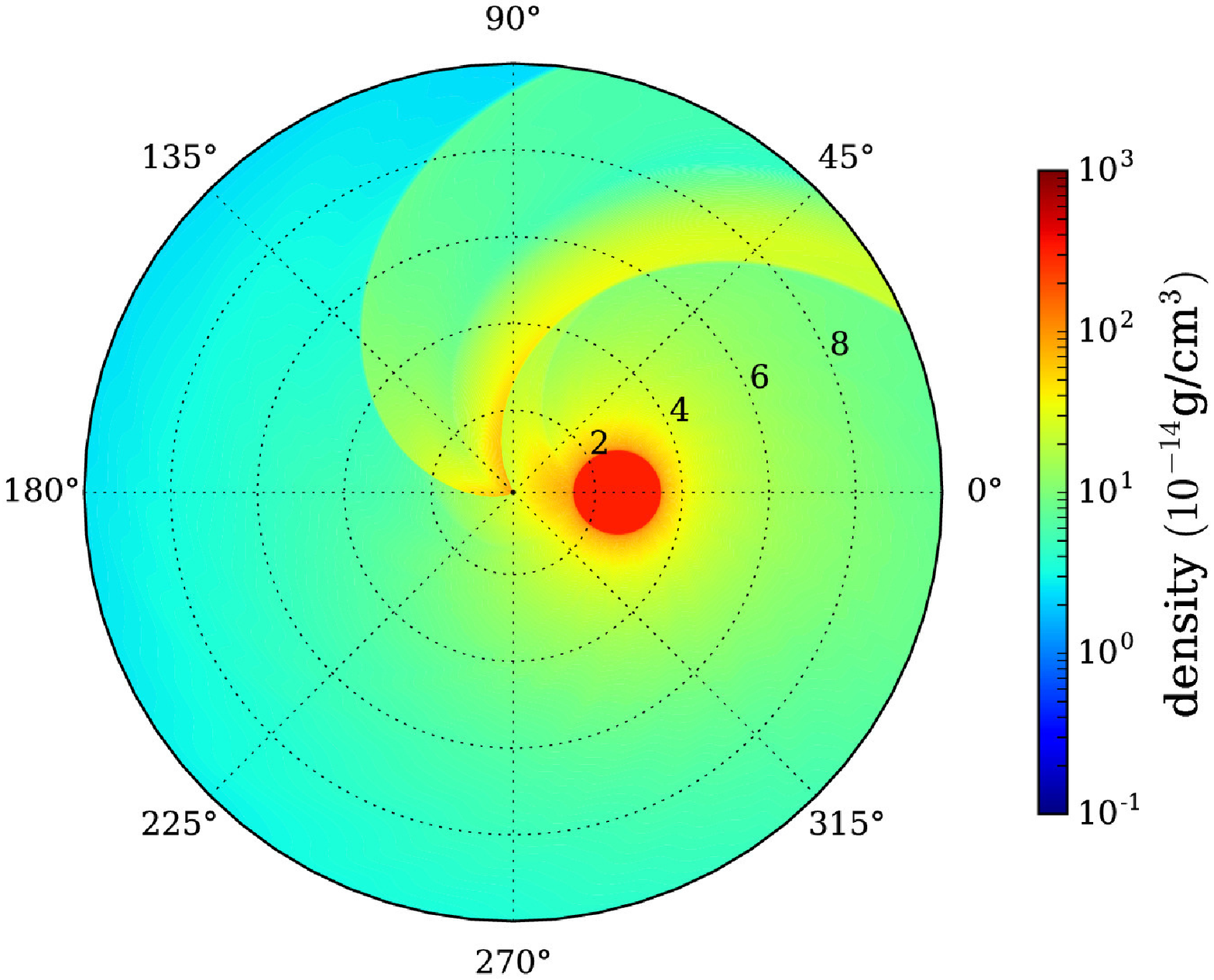}{0.5\textwidth}{(d)}
  }
  \caption{
The same as Figure \ref{fig:1} except for LMC X-1.
In the panel (a) and (b), the positions of BH and the surface of companion star are located at the origin and $1.5 \times 10^{12}$ cm respectively.
 } 
 \label{fig:3}
   \end{center}
\end{figure*}

We present the results of Cyg X-1 in HS in Figure \ref{fig:1}, and LH state in Figure \ref{fig:2}, respectively. 
Figure \ref{fig:3} shows the results of LMC X-1.
The density is plotted in each panel (a) in blue points. 
A continuous spherical wind model (= $\rho=\dot{M}_* /(4\pi r^2 V_{\rm{empirical}})$) is overlaid in a black curve. 
A cylindrical wind model (= $\rho \propto r^{-1} V_{\rm{empirical}}^{-1}$) is shown in a red curve. 
The density close to the BH (at radius $\sim$ 0) is about an order of magnitude higher than that of the spherical wind model, 
and a little higher than that of the cylindrical wind model. 
This is simply because the BH gravity focuses the wind. 
The panel (b) shows the velocity dependence on radius. 
The velocity distribution brakes at the He$\rm{{I\hspace{-.1em}I}}$ ionization front as indicated in a green vertical dotted-line. 
For the LH state (in Figure \ref{fig:2}b), we cannot clearly see the break 
since it is located near the area where the BH gravity becomes comparable to the wind acceleration 
(i.e., the wind velocity there is not solely controlled by wind acceleration, but the BH gravity).


\begin{deluxetable*}{cccccc}[h!]
\tablecaption{The typical values in our HD calculation results.\label{tab_v4:3}}
\tablecolumns{6}
\tablewidth{0pt}
\tablehead{
 \colhead{Source name}&\colhead{He$\rm{{I\hspace{-.1em}I}}$ front [cm]}&\colhead{$V_{\rm{wind}}\tablenotemark{a}$ [cm s$^{-1}$]}&\colhead{$R_{\rm{acc}}$ [cm]}&\colhead{ $\rho_{\rm{acc}}\tablenotemark{b}$ [g cm$^{-3}$]}&\colhead{$\dot{M}_{\rm{cap}}$ [g s$^{-1}$]}
}
\startdata
 Cyg X-1 (HS)&  $1.94-1.96\times10^{12}$&$0.51-0.57\times10^{8}$&$1.2-1.5\times 10^{12}$&$3\times10^{-14}$&$0.8-1\times10^{19}$\\ 
Cyg X-1 (LH)&$9.21\times10^{11}$&$1.3\times10^{8}$&$ 0.23\times10^{12}$&$2\times10^{-14}$&$5\times10^{17}$ \\ 
 LMC X-1&$1.46\times10^{12}$&$0.53\times10^{8}$&$0.95\times 10^{12}$&$3\times10^{-13}$&$5\times10^{19}$\\ 
\enddata
\tablenotetext{a}{Radial velocity at the He$\rm{{I\hspace{-.1em}I}}$ ionization front and $\varphi=0$.}
\tablenotetext{b}{Density at BH and $\varphi=0$.}
\tablecomments{Radius of He$\rm{{I\hspace{-.1em}I}}$ ionization front fluctuates by a few percent for HS state in Cyg X-1.}
\end{deluxetable*}

In Table \ref{tab_v4:3}, we list the typical values of the physical quantities obtained by the HD calculations; He$\rm{{I\hspace{-.1em}I}}$ ionization front, wind velocity, accretion radius, density and captured mass rate, respectively. 
Here, we define the captured mass rate
 \begin{equation}
\dot{M}_{\rm{cap}} = \pi R_{\rm{acc}}^2\rho_{\rm{acc}} V_{\rm{wind}},
\label{eq:a}
\end{equation}
where $R_{\rm{acc}}=2GM_X/V_{\rm{wind}}^2$ is accretion radius and $V_{\rm{wind}}$ denotes the velocity at the He$\rm{{I\hspace{-.1em}I}}$ ionization front and $\varphi=0$.
$\rho_{\rm{acc}}$ is the density at the BH and $\varphi=0$.
The front is updated at every time step using Eq. (\ref{eq:1}) for HS state and Eq. (\ref{eq:2}) for LH state, 
which makes the front fluctuate between two cells when the cell size is not fine enough to resolve the exact position of the front. 
In fact, for HS state of Cyg X-1, the radius of the front varies by a few percent $R_{\rm{S}}=1.94\times 10^{12}-1.96\times 10^{12}$ cm as shown in Table \ref{tab_v4:3}.
It took more often the smaller value $R_{\rm{S}}=1.94\times 10^{12}$ cm, so that we adopt the smaller as the representative value of Cyg X-1 HS state for analytical calculations; accordingly, it corresponds to the higher wind velocity $V_{\rm{wind}}=0.57\times 10^8$ cm s$^{-1}$. 
For LH state in Cyg X-1 and LMC X-1, the fluctuation of the front was negligibly small (less than one percent).
We also investigated the wind-fed process in the rotational frame by using the default function implemented in \software{PLUTO}. 
The density contours are presented in the panels (c) without rotation and (d) with rotation.  
Although the stream lines were changed by binary rotation, the typical values of density and velocity differed only by a few percent. 
Since our discussion utilizes the values at the He$\rm{{I\hspace{-.1em}I}}$ front with $\varphi=0$ and close to the BH, 
we do not proceed to further details on the rotational effects.

If the simulation results are taken as the face values, 
the captured mass rate would become higher than the mass accretion rate expected from the observed X-ray luminosity; 
especially in the HS state, e.g., $\dot{M}_{\rm{cap}} \gg \dot{M}_{\rm{expected}}\sim L_X/(0.1c^2)\sim10^{17}$ g s$^{-1}$ for HS state of Cyg X-1 and $\dot{M}_{\rm{cap}} \gg \dot{M}_{\rm{expected}}\sim10^{18}$ g s$^{-1}$ for LMC X-1. 
To fill the gap between the rates, we suggest that there are some outflowing and stagnating gasses to control the net accretion rate,  
which have a high temperature by heating via shock or X-ray irradiation from the BH. 
The obtained optical depth for electron scattering is quite small, $\tau \sim 2l\rho\kappa_{\rm{es}} \sim R_{\rm{acc}}\rho_{\rm{acc}} \ll 1$, where $l$ is the scale length and $ \kappa_{\rm{es}}$ represents electron scattering opacity. 
Since the captured gas is optically thin along $l\sim R_{\rm{acc}}$, the radiation could not be blackbody emission. 
It prevents us from explaining the huge gap between $\dot{M}_{\rm{cap}}$ and $\dot{M}_{\rm{expected}}$ 
by a simple scenario that the captured gas is cooled enough to turn into a standard disk. 
We thus consider that an optically thin and geometrically thick flow is formed at the outside of the standard disk; 
hereafter it is called ``torus-like flow'', or ``black-hole binary torus''.

\subsection{Magneto-rotational instability (MRI)}\label{sec:2.2}

Magneto-Rotational Instability (MRI) is one of the most probable mechanisms for extraction of angular momentum from accreting gas.
The fundamental process is triggered by magnetic tensions acting on charged particles in differentially rotating disks.
This idea is suggested by Balbus \& Hawley (1991) and Hawley \& Balbus (1991). 
In the following, we present that MRI in the torus-like flow is important physical process to connect the captured wind with the accretion flow.

When the MRI on-set condition
\begin{equation}
v_{\rm{A\phi}}^2<c_{\rm{s}} v_{\rm{K}}, 
\label{eq:10}
\end{equation}
is satisfied in the torus-like flow, it is expected that a steady accretion is established inside a critical radius where $v_{\rm{K}}$ is the Keplerian speed (Pessah \& Psaltis 2005; Begelman \& Pringle 2007; Begelman et al. 2015).
In brief, this equation means that the time scale for stabilizing the fluctuations of the line of magnetic force 
is longer than the one for establishing local pressure balance and the rotation of the flow.
Although Eq. (\ref{eq:10}) is helpful for analytical investigation, it is uncertain whether the MRI is completely stabilized above the threshold.
We consider that the toroidal magnetic pressure is responsible for determination of the scale height of the flow.
Begelman et al. (2015) showed that the toroidal magnetic pressure is roughly independent of vertical direction $z$ in the MRI on-set layer.
Thus, we ignore the effect of the vertical structure of the magnetic field and write $B_{\phi}$ and $v_{\rm{A\phi}}$ just as $B$ and $v_{\rm{A}}$.
It should be noted that there is also a poloidal magnetic field and it would be needed to keep the magnetically dominated flow (Salvesen et al. 2016; Fragile and Sadowski 2017).
In the present paper, we suppose that the poloidal magnetic pressure is always weaker than the toroidal one, and does not affect directly the flow dynamics.

Here, we put the important assumption that the magnetic field lines are transported from the companion. 
This idea is supported by optical polarization measurement that the magnetic field around the outer rim of the accretion disk of Cyg X-1 reaches about a few hundred Gauss (Karitskaya et al. 2010). 
The right-hand side of Eq. (\ref{eq:10}) scales as $c_{\rm{s}}v_{\rm{K}}\propto T^{1/2} R^{-1/2}$ for ideal fluid, while the left-hand side scales as $v_{\rm{A}}^2 \propto B^2/\rho$.
Thus, we consider that the MRI in the torus-like flow does not operate until the density reaches a critical value.
In other words, the MRI on-set condition works like a water gate for the total accreting matter.

Using Eq. (\ref{eq:10}), the critical density is expressed as 
\begin{equation}
\rho_c=\frac{B^2}{4\pi}c_{\rm{s}}^{-1} v_{\rm{K}}^{-1}|_{R=R_c},
\label{eq:13}
\end{equation}
where $R_c$ denotes the radial distance of the torus-like flow from the BH at the critical condition.
The net accretion rate is written as
\begin{equation}
\dot{M}_{\rm{net}}=4\pi RH \rho v_{\rm{in}}|_{R=R_c},
\label{eq:14}
\end{equation}
where $H$ is the scale height of the torus-like flow at the MRI on-set condition (i.e., the magnetic pressure is dominant)
and $v_{\rm{in}}$ is the radial velocity, 
\begin{equation}
\frac{H}{R}\sim \frac{v_A}{v_{\rm{K}}}\sim \left(\frac{c_{\rm{s}}}{v_{\rm{K}}}\right)^{1/2},
\label{eq:15}
\end{equation}
\begin{equation}
v_{\rm{in}}\sim \frac{3}{2} \alpha \left(\frac{H}{R}\right)^2 v_{\rm{K}} \sim \frac{3}{2} \alpha c_{\rm{s}},
\label{eq:16}
\end{equation}
where $\alpha$ is viscous parameter that relates to viscosity as $\nu_{\rm{visc}}=\alpha v_{\rm{A}}H$ (Begelman \& Pringle 2007). 
Moreover, we have
 \begin{equation}
 \dot{M}_{\rm{net}} = L_{X}/ \epsilon c^2.
 \label{eq:17}
 \end{equation}
where $\epsilon$ is radiative efficiency. 
We assume $\alpha = 0.15$ and $\epsilon=0.1$ through out the paper. 
 
Here, we introduce a parameter $\zeta$ which relates the accretion radius $R_{\rm{acc}}$ to the critical radius $R_c$
so that 
\begin{equation}
R_c=\zeta R_{\rm{acc}}=\frac{2GM_X}{c_{\rm{s,shock}}^2},
\label{eq_v3:8}
\end{equation}
where $\zeta \equiv V_{\rm{wind}}^2/c_{\rm{s, shock}}^2$, $c_{\rm{s, shock}}=\sqrt{\frac{k_BT_{\rm{sh}}}{\mu m_p}}$ is sound speed of the adiabatic strong shocked gas, $T_{\rm{sh}}$ is shock temperature and $\mu\sim0.5$ is mean molecular weight.
It means that the scale length of the torus-like flow is determined by the sound speed of shocked matter, rather than the wind velocity at the He$\rm{{I\hspace{-.1em}I}}$ ionization front.
We expect that $c_{\rm{s, shock}}$ does not depend on $V_{\rm{wind}}$.
This is because the gravity determines the speed of the captured gas near the BH.

We estimate very roughly the fiducial value $\zeta_{\rm{f}}$.
The shock temperature is written by Keplerian speed, 
\begin{equation}
k_BT_{\rm{sh}}\approx \frac{3m_p}{16}\times (2v_{\rm{K}}(R))^2,
\label{eq:12}
\end{equation}
where the factor of two comes from the fact that kinetic energy of the gas is dissipated through head-on collision.  
In order to determine the shock temperature, a certain criterion of the radius is necessary. 
Then, we define the effective accretion radius by taking the shock heating into account as 
 \begin{equation}
R_{\rm{acc,eff}} =\frac{2GM_X}{V_{\rm{wind}}^2+c_{\rm{s,shock}}^2}.
\label{eq:11}
\end{equation}
Substituting  Eq. (\ref{eq:11}) for Eq. (\ref{eq:12}) and replacing $T_{\rm{sh}}$ with $c_{\rm{s, shock}}$,
\begin{equation}
\mu c_{\rm{s, shock}}^2 \approx \frac{3}{8}(V_{\rm{wind}}^2 + c_{\rm{s, shock}}^2),
 \label{eq:v8_1}
\end{equation}
and then, we obtain a fiducial value of $\zeta$ as 
\begin{equation}
\zeta_{\rm{f}} \approx1/3.
 \label{eq:v8_2}
\end{equation}
Note that the value of $\zeta$ can change with the velocity so that $\zeta_{\rm{f}}$ means just a certain criterion.
Using the results of calculations, we derived $\zeta$ $( \propto V_{\rm{wind}}^2)$ by adjusting the mean of the two values to become $\approx 1/3$ for Cyg X-1, while $\zeta = \zeta_{\rm{f}}$ for LMC X-1 as in Table 4.
The effect of this assumption is discussed later.

Next, we investigate the temperature $T$ and the magnetic field strength parameter $B$ of the torus-like flow at $R_c$.
We expect that the captured gas cools with increasing density, and the temperature deviates from the shock temperature $T_{\rm{sh}}$.
For simplicity we first adopt $T_c = 10^6$ K, and then we check the ionization parameter and the self-consistency of this assumption.
We calculate the net accretion rate from Eq. (\ref{eq:14})-(\ref{eq:17}), and solve it for the critical density to be consistent with X-ray luminosity (Table \ref{tab_v4:1}).
After that, we have obtained the ionization parameter at $R_c$,
\begin{equation}
\xi_c= \frac{m_p L_X}{ \rho_cR_c^2}\sim 10^3 \mbox{ erg cm s$^{-1}$}, 
 \label{eq:18}
 \end{equation}
 for all calculations (Table \ref{tab_v4:4}). 
In such highly ionized gas, line cooling is inefficient and the characteristic temperature becomes $\sim 10^6$ K (Nakayama \& Masai 2001).
Because the captured gas has a large vertical scale height, this estimation seems to be consistent with a geometrically-thick flow irradiated by X-rays from the BH.
Therefore, in the following, we adopt $T_c =10^6$ K for all estimations. 
Moreover, from Eq. (\ref{eq:13}), we estimate the value of magnetic field strength parameter at $R_c$ 
\begin{equation}
B_c=\sqrt{4\pi \rho_c c_{\rm{s}}v_{\rm{K}}} = 112 \times \left(\frac{\rho_c}{10^{-12} \mbox{ g cm$^{-3}$}}  \right)^{1/2} \left(\frac{c_{\rm{s}}}{10^{7} \mbox{ cm s$^{-1}$}}  \right)^{1/2} \left(\frac{v_{\rm{K}}}{10^{8} \mbox{ cm s$^{-1}$}}  \right)^{1/2} [\mbox{G}].
 \label{eq_v5:1}
 \end{equation}
The results are listed in Table \ref{tab_v4:4}.
The values of $B_c \approx 10^2$ G are comparable to or higher than the O-type star observations (Bychkov et al. 2009; Karitskaya et al. 2010).
Because the density of the captured wind increases up to the critical value, we expect that the energy density of the magnetic field must also increase.
The ratio $P_{\rm{mag}}/P_{\rm{gas}}$ at $R_c$ is $\sim 3-5$, and becomes smaller outside $R_c$ where $P_{\rm{mag}}$ and $P_{\rm{gas}}$ are the gas and magnetic pressures.

In our model, because the companion of LMC X-1 has a higher mass loss rate $\dot{M}_*$ than that of Cyg X-1, 
it is expected that a higher total magnetic field energy is transported from the companion to the torus-like flow in LMC X-1.
This is the reason for the difference between LMC X-1 and Cyg X-1.
Note that we adopted the fiducial value of $\zeta=1/3$ for LMC X-1, 
but even if we purposely take $\zeta=0.11$ which is the same as Cyg X-1 HS state, 
the estimations almost do not change; the ionization parameter becomes $\sim1.2\times 10^3$ erg cm s$^{-1}$, and the magnetic field strength parameter $\sim 7.3\times 10^2$ G.

The luminosity of the torus-like flow is estimated to be less than one percent of the total X-ray luminosity $L_X$.
However, narrow recombination continua and iron K-lines, which are emitted in the region of $kT \sim 100$ eV around $\log{\xi_c} \sim 3$ (Nakayama \& Masai 2001), may be detected from the torus-like flow if observations are capable of high-resolution spectroscopy with good statistics. 

\begin{deluxetable*}{ccccccc}[h!]
\tablecaption{Wind velocities and other parameters.\label{tab_v4:4}}
\tablecolumns{6}
\tablewidth{0pt}
\tablehead{
 \colhead{Source name}&\colhead{$V_{\rm{wind}}$ [cm s$^{-1}$]}&\colhead{$\zeta$}&\colhead{$R_c$\tablenotemark{a} [cm]}&\colhead{$\rho_c$ [g cm$^{-3}$]}&\colhead{$B_c$ [G]}&\colhead{$\xi_c$ [erg cm s$^{-1}$]}
}
\startdata
 Cyg X-1 (HS)&$0.57\times 10^8$&0.11&$1.3\times10^{11}$&$3.0\times10^{-12}$&$2.4\times 10^2$&$1.1\times 10^3$\\ 
Cyg X-1 (LH)&$1.3\times 10^8$&0.56&$1.3\times10^{11}$&$0.75\times10^{-12}$&$1.2\times10^2$&$1.1\times10^3$\\ 
 LMC X-1&$0.53\times 10^8$&0.33&$3.2\times10^{11}$&$2.5\times10^{-12}$&$1.6\times10^2$&$1.6\times10^3$\\ 
\enddata
\tablenotetext{a}{Critical radius where the MRI begins to operate in the torus-like flow.}
\tablecomments{The ionization parameters of all cases are high enough to be $T_c\sim10^6$ K.}
\end{deluxetable*}

Now we examine the accretion process of the torus-like flow. 
Increasing the density up to the critical value $\rho_c$ and having the high ionization parameter $\log{\xi_c} \sim 3$ suggest that the bremsstrahlung cooling dominates and the line cooling is negligible. 
We can write a ratio between the cooling rate of bremsstrahlung and the release rate of the gravitational energy as
\begin{equation}
\frac{RH \rho^2 T^{1/2}} {\dot{M}_{\rm{net}}/R^2} \propto \rho R^2 \propto L_X/\xi, 
\label{eq:20}
\end{equation}
here, we used
\begin{equation}
\dot{M}_{\rm{net}} = 4\pi RH\rho v_{\rm{in}} = \mbox{constant}.
\label{eq:21}
\end{equation}
We assume the cooling-rate vs. gravitational-energy ratio do not change in a spectral state, but can change if a state transition occurs. 
It means the variation of X-ray luminosity in a state can directly affect the ionization parameter at each radius. 
For another interpretation, the more cooling becomes effective, the greater amount of gas can accrete. 
 From Eq. (\ref{eq:15}), (\ref{eq:16}), (\ref{eq:20}) and (\ref{eq:21}), we obtain the scaling relations of the accreting torus-like flow
\begin{equation}
\rho \propto R^{-2},~~\\
T \propto R^{-1/3},~~\\
H/R \propto R^{1/6}. \\
\label{eq:22}
\end{equation}

As mentioned at the beginning of section \ref{sec:2}, we consider that the multi-color disk begins to form where $\tau \gtrsim 1$.
Using Eq. (\ref{eq:22}), the optical depth scales as 
\begin{equation}
\tau = 2\rho H \kappa_{\rm{es}} \propto R^{-5/6}.
\label{eq:23}
\end{equation} 
We calculate radius $R_d$, density $\rho_d$ and temperature $T_d$ where the optical depth becomes $\tau=1$.
The results are listed in Table \ref{tab_v4:5}.
For Cyg X-1, the radii $R_d$ in both states become around an order of magnitude smaller than LMC X-1's which is comparable to a typical disk size of LMXBs.
The temperatures $T_d$ are relatively higher than that of the multi-color disk.
However, when the flow becomes optically thick, the blackbody radiation dominates and immediately cools the gas.
We find that these results do not depend so much on the parameter $\zeta$ because $R_d\propto R_c^{-1/5}$.

\begin{figure*}[b]
 \begin{center}
\includegraphics[scale=0.7]{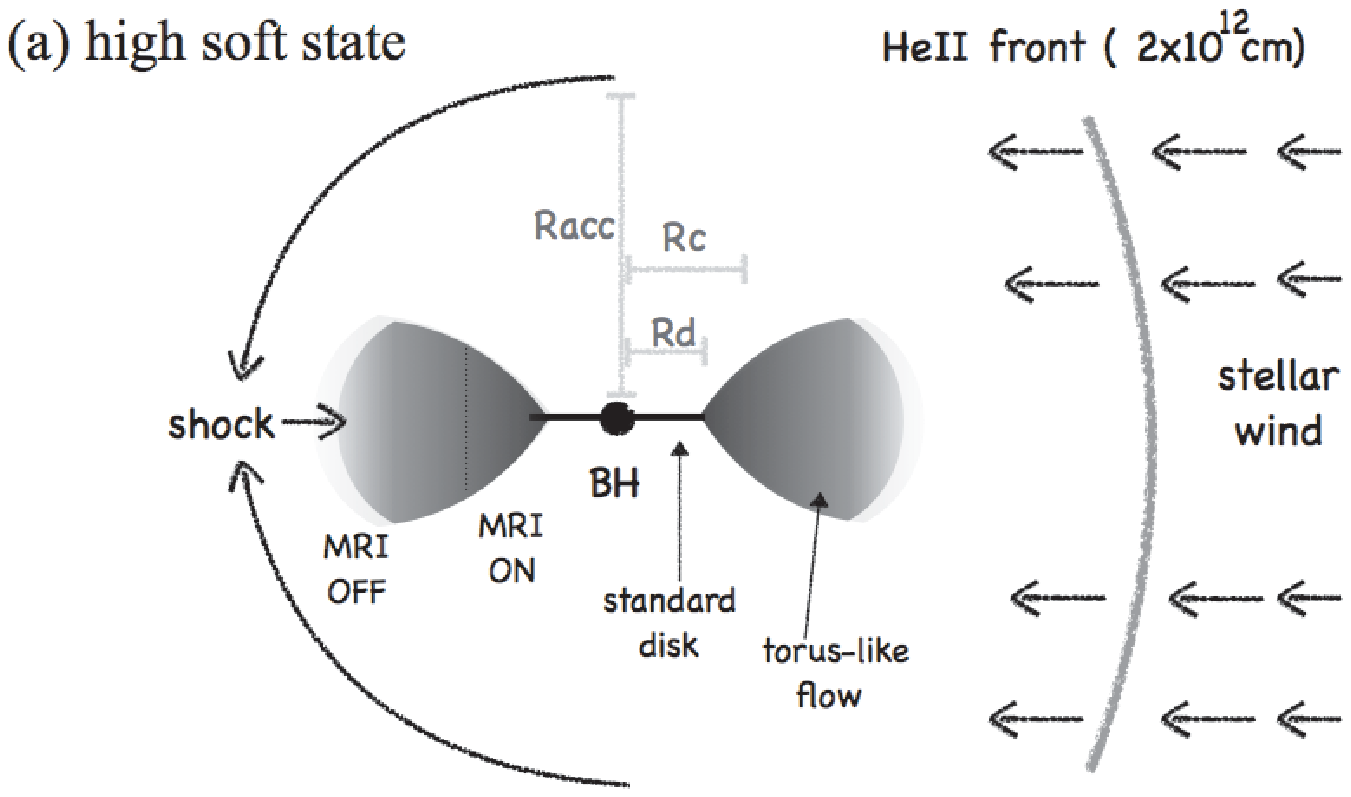}
\includegraphics[scale=0.7]{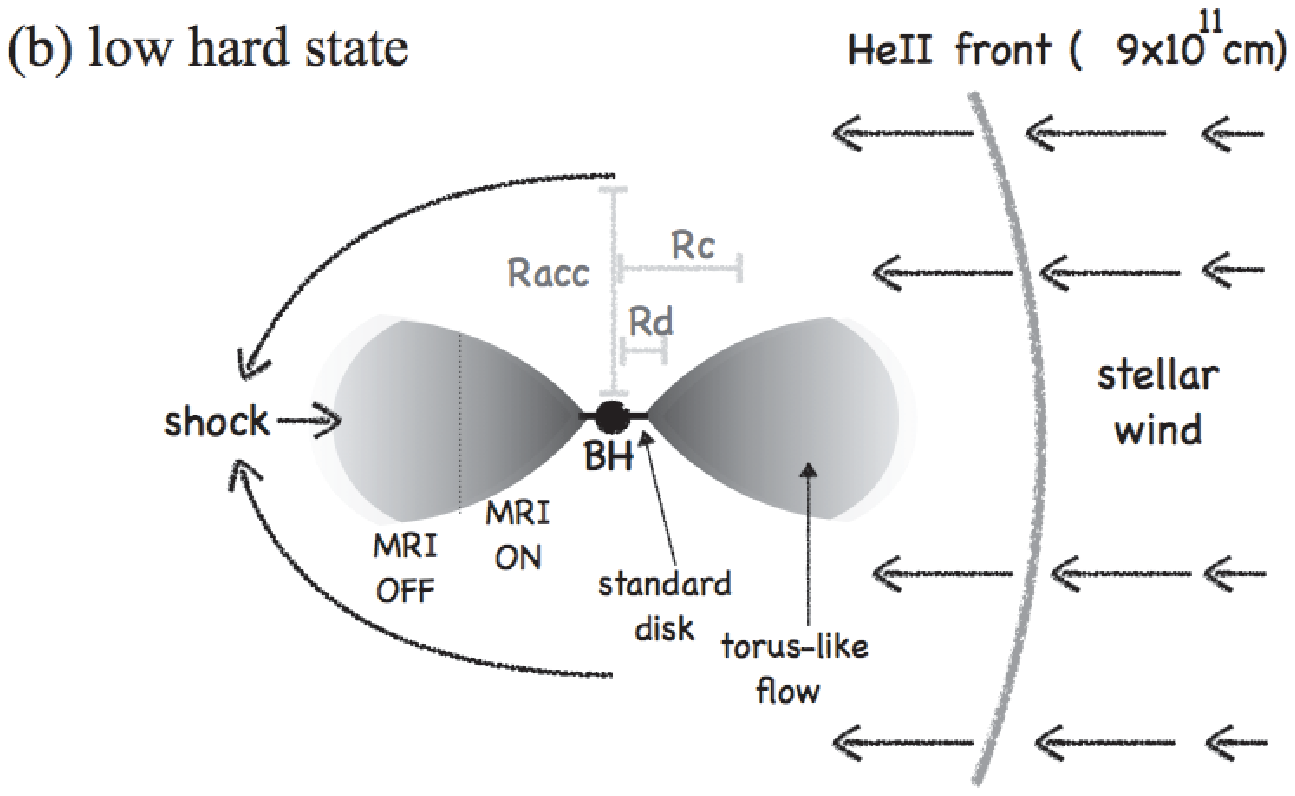}
 \caption{
Schematic illustrations of the present model: (a) HS and (b) LH states.
The arrows represent the flow of the accreting matters.
The acceleration of line-driven wind terminates at the He$\rm{{I\hspace{-.1em}I}}$ front.
The gradation of torus-like flow expresses the density.
The flow becomes standard thin disk-like well inside $R_d$.
}
 \label{fig:4}
\end{center}
\end{figure*}

\begin{figure*}[h!]
 \begin{center}
 \plotone{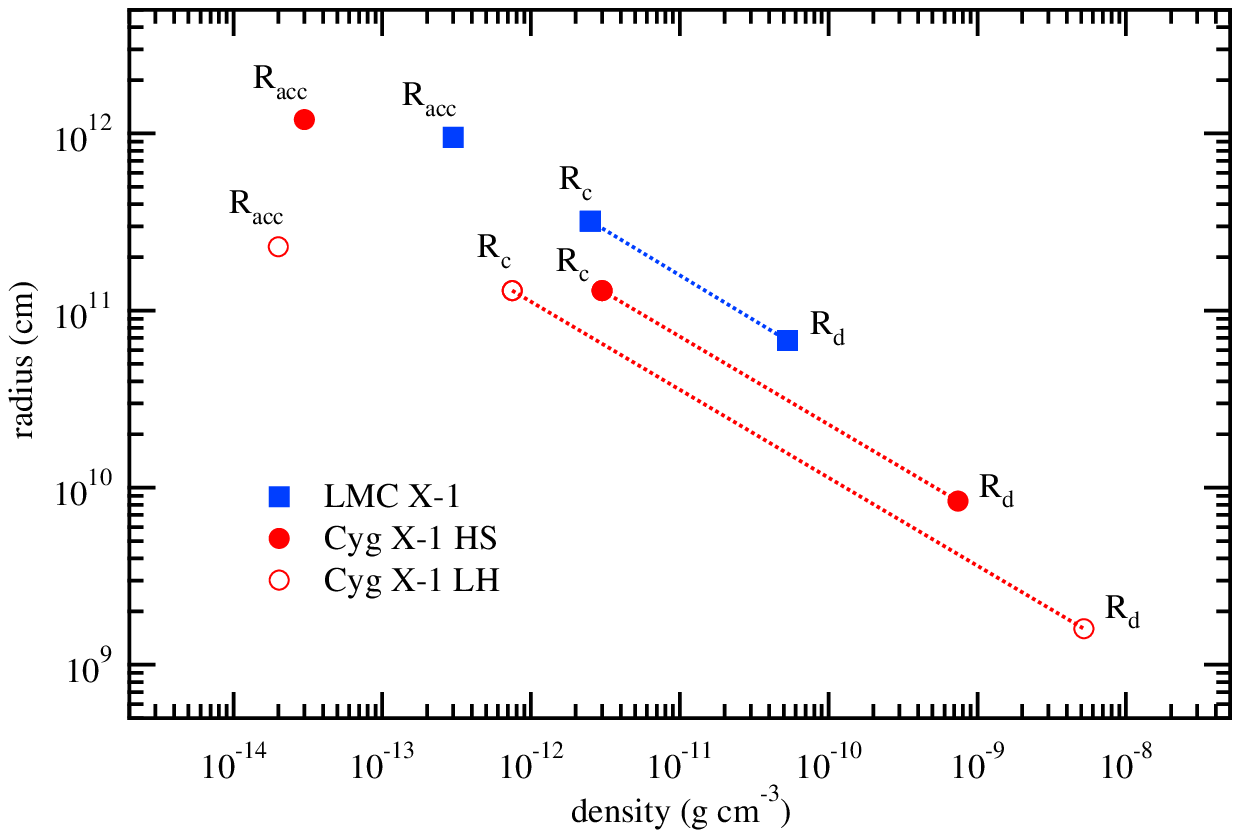}
  \caption{
$R_{\rm{acc}}$, $R_c$ and $R_d$, vs. $\rho_{\rm{acc}}$, $\rho_c$ and $\rho_d$ for Cyg X-1 HS (red filled circles) and Cyg X-1 LH states (red open circles), and LMC X-1 HS state (blue filled boxes). }
 \label{fig:5}
\end{center}
\end{figure*}

Figure \ref{fig:4} shows schematic illustrations of the present model for HS and LH states.
The radius of the He$\rm{{I\hspace{-.1em}I}}$ front varies with the X-ray spectrum and affects $V_{\rm{wind}}$ and $R_{\rm{acc}}$.
The shocked wind matter forms a torus-like flow and it cools with increasing density.
At $R_c$ where the density reaches a critical value $\rho_c$, the MRI begins to operate and the net mass accretion rate is determined.
Since the stellar wind continuously supplies the gas to the torus-like flow, steady accretion would be established.
The flow becomes standard thin disk-like well inside $R_d$.
As a result, three scale lengths are proposed to describe the accretion physics at the outer region of accretion disks in BH HMXBs.
To illustrate the results, we plotted the various radii (i.e. $R_{\rm{acc}}$, $R_c$ and $R_d$) and densities ($\rho_{\rm{acc}}$, $\rho_c$ and $\rho_d$) in Figure \ref{fig:5}.
 
It should be noted that the accretion flow well inside $R_d$ is thought not to differ from that in LMXBs.
Therefore, we expect that the magnetic field from the companion could act to stabilize thermal instability of the standard thin disk (Begelman \& Pringle 2007; Sadowski 2016).

 \begin{deluxetable}{ccccc}[h!]
\tablecaption{The physical quantities at $R_c$ and $R_d$ are listed. $T_c=10^6$ K is assumed.\label{tab_v4:5}}
\tablecolumns{6}
\tablewidth{0pt}
\tablehead{
 \colhead{Source name}&\colhead{$\tau_c$\tablenotemark{a}}&\colhead{$R_d$\tablenotemark{b} [cm]}&\colhead{$\rho_d$ [g cm$^{-3}$]}&\colhead{ $T_d$ [K]}
}
\startdata
Cyg X-1 (HS)&0.10&$8.4\times10^{9}$&$7.4\times10^{-10}$&$2.5\times10^{6}$\\ 
Cyg X-1 (LH)&0.025&$1.6\times10^{9}$&$5.2\times10^{-9}$&$4.3\times10^{6}$\\ 
LMC X-1&0.28&$6.8\times10^{10}$&$5.3\times10^{-11}$&$1.7\times10^{6}$\\ 
\enddata
\tablenotetext{a}{Critical optical depth where the torus-like flow begins to accrete via MRI.}
\tablenotetext{b}{Radius where the optical depth becomes unity.}
\end{deluxetable}

\section{discussion}\label{sec:3}
Here, we discuss the state transition mechanism using the results of our calculations and the idea of MRI determining the net accretion rate.  
We focus on the questions about the transitions; why Cyg X-1 has two relatively stable states and undergoes transitions, while a transition has not been observed in LMC X-1.
A remarkable feature between the two objects is the huge difference in luminosity though they show similar time variability. 
The luminosity of Cyg X-1 is typically about $\sim 10^{37}$ erg s$^{-1}$ (Gierli\'{n}ski et al. 1999; Zdziarski et al. 2002), while that of LMC X-1 is $\sim10^{38}$ erg s$^{-1}$ (Long, Helfand \& Grabelsky 1981; Gou et al. 2009).
If LMC X-1 transits to the LH state at the same luminosity as Cyg X-1, its luminosity needs to decrease by a factor of $\sim$ 10; 
i.e.,  $\Delta L_X/L_X\sim10$ is necessary to cause the state transition. 
It is larger than the observed variability $\Delta L_X/L_X < 4$ (Ruhlen et al. 2011). 
We infer that this is the reason why LMC X-1 has not shown a transition. 
Thus, we will try to present a mechanism for the variability of the mass accretion rate, which would directly/indirectly relate with the state transition. 
In addition, we discuss why the stable states are kept maintained for a long time. 

\subsection{trigger of state transitions}\label{sec:3.1}

Because Cyg X-1 and LMC X-1 show a similar level of $\Delta L_X$, 
we infer that the variability is caused by the same mechanism in the two.
From Eq. (\ref{eq:13})- Eq. (\ref{eq_v3:8}), the X-ray luminosity scales as
\begin{equation}
 L_X\propto \dot{M}_{\rm{net}} \propto B_c^2 T_c^{1/4}.
\label{eq_v3:11}
\end{equation} 
 Since the dependence on the temperature is weak, it cannot be crucial.
 Thus, we consider the fluctuation of the critical magnetic field strength $\delta B_c (\lesssim B_c)$ as a main cause of the variability.
  The fluctuation $\delta B_c(\lesssim B_c)$ can make the changes in the net accretion rate and the X-ray luminosity by $\Delta \dot{M}_{\rm{net}}/\dot{M}_{\rm{net}} \sim \Delta L_X/L_X \sim 4$.
  The factor of four is based on the observations (Nowak et al. 2012; Zdziarski et al. 2002; Wilms et al. 2001; Ruhlen et al. 2011). 
  Moreover, this mechanism can also explain the fact that LMC X-1 in LH state has never been observed.  
  $\delta B_c$ cannot be larger than $B_c$ unless the magnetic field structure changes drastically. 
  Therefore, LMC X-1  does not switch state.
 Note that the fluctuation $\delta B_c$ introduced here is different from the field amplified with decreasing radius due to the MRI.
Instead it represents the fluctuation of initial field strength at $R_c$ where the MRI sets in.
 
In Section \ref{sec:2.1}, we mentioned that the radius of the He$\rm{{I\hspace{-.1em}I}}$ ionization front determines the terminal wind velocity, and is changed by variations of the X-ray luminosity and the spectrum as Eq. (\ref{eq:1}) and Eq. (\ref{eq:2}). 
 From this point of view, the wind velocity does not become a trigger for state transitions, but makes a difference in the features of the two states as the result of transition.
In order to take into account that the change in wind velocity affects the wind density and the energy density of magnetic field, we suppose that the magnetic field strength is related as
\begin{equation}
B_c^2 \propto \rho_c \propto \rho_{\rm{wind}} \propto V_{\rm{wind}}^{-1} r_{\rm{wind}}^{-p},
\label{eq_v3:9}
\end{equation} 
where $r_{\rm{wind}}$ is the distance from the OB-type star to the He$\rm{{I\hspace{-.1em}I}}$ ionization front with $\varphi=0$, $\rho_{\rm{wind}}$ is wind density at $r_{\rm{wind}}$ and $p$ is a geometrical parameter ($p=1$ means cylindrical wind). 
We expect that $p$ is independent of the spectral states.
By equating $p$ in the two states, we obtain 
\begin{equation}
p=\log_{r_{\rm{wind,LH}}/r_{\rm{wind,HS}}}\left(\frac{\rho_{\rm{wind,HS}} V_{\rm{wind,HS}}}{\rho_{\rm{wind,LH}} V_{\rm{wind,LH}}} \right) \sim0.79, 
\label{eq_v3:10}
\end{equation} 
where the subscripts refer to the two spectral states. 
Combining Eq. ({\ref{eq_v3:11}}) and Eq. ({\ref{eq_v3:9}}), we obtain
\begin{equation}
L_X\propto \dot{M}_{\rm{net}}\propto V_{\rm{wind}}^{-1}r_{\rm{wind}}^{-p}.
\label{eq_v3:12}
\end{equation}  
A relatively large fluctuation $\delta B_c(\lesssim B_c)$ could change the X-ray spectrum as well as the X-ray luminosity.
The significant change in the X-ray spectrum leads to variations of both $V_{\rm{wind}}$ and  the radius of the He$\rm{{I\hspace{-.1em}I}}$ ionization front (i.e. $r_{\rm{wind}}$).
The variations can make the change in $B_c$ by a factor of two (Table \ref{tab_v4:3} and Table \ref{tab_v4:4}).

Even if the magnetic field strength has a very small fluctuation, the luminosity and the wind velocity could be varied.
However, using Eq. ({\ref{eq:1}}) and Eq. ({\ref{eq:2}}), we obtain
\begin{equation}
R_{\rm{S}}\propto \rho_{\rm{wind}}^{-2/3}L_X^{1/3} \propto V_{\rm{wind}}^{1/3}r_{\rm{wind}}^{p/3} \propto B_c^{-2/3},
\label{eq_v3:13}
\end{equation}  
\begin{equation}
R_{\rm{He}}\propto \rho_{\rm{wind}}^{-3/5}L_X^{2/5}\propto V_{\rm{wind}}^{1/5}r_{\rm{wind}}^{p/5}  \propto B_c^{-2/5}.
\label{eq_v3:14}
\end{equation}  
These relations show negative feedback of $B_c$ to the He$\rm{{I\hspace{-.1em}I}}$ ionization front and the initial small change in $B_c$ does not affect the radius of the front unless the X-ray spectrum changes sufficiently.
Therefore, our model can naturally explain why Cyg X-1 stays in a state for a relatively long time ($\sim 10-1000$ days) without any periodic features (Wen et al. 2001; Grinberg et al. 2013; Grinberg et al. 2014; Sugimoto et al. 2016).
Moreover, from the aspect of stability, we conclude that because the He$\rm{{I\hspace{-.1em}I}}$ ionization front does not move drastically, LMC X-1 does not show state transitions.
 We note that this stability is due to the assumption of Eq. (\ref{eq_v3:8}).
Instead, if $R_c$ is determined by the accretion radius $R_{\rm{acc}}\propto V_{\rm{wind}}^{-2}$, Eq. (\ref{eq_v3:11}) is replaced by 
\begin{equation}
L_X\propto \dot{M}_{\rm{net}}\propto B^2 V_{\rm{wind}}^{-11/2}T^{1/4}.
\label{eq_v3:15}
\end{equation}  
This is unstable for luminosity variations due to fluctuations of the magnetic field strength.

In the HD calculations of Cyg X-1, we performed iterative computations to determine the parameter $\dot{M}_*$:
we choose the parameter $\dot{M}_*$, obtain the velocity and density of each state from HD calculations, and then compute the ratio
\begin{equation}
\frac{L_{X,\rm{HS}}}{L_{X,\rm{LH}}}\sim \frac{\dot{M}_{\rm{net,\rm{HS}}}}{\dot{M}_{\rm{net,\rm{LH}}}}\sim \left(\frac{V_{\rm{wind,HS}}}{V_{\rm{wind,LH}}}\right)^{-1} \left(\frac{r_{\rm{wind,HS}}}{r_{\rm{wind,LH}}}\right)^{-p}\sim \left(\frac{\rho_{\rm{wind,HS}}}{\rho_{\rm{wind,LH}}}\right).
\label{eq:b}
\end{equation} 
The simulation results of Cyg X-1 presented in this paper were derived by computing
iteratively until this ratio becomes a factor of $\sim 4$ (The factor of $\sim 4$ is a observational fact shown in Table \ref{tab_v4:1}).
Therefore, we succeeded in obtaining a consistent solution in that the
simulation results are what we expect from our arguments based on Eq. (\ref{eq_v3:11}) - Eq. (\ref{eq:b}).

 \subsection{accretion rate in MRI onset layers}\label{sec:3.2}

In our model, we assume that the state transition could occur in Cyg X-1 with only a factor of $\lesssim 4$ increase (decrease) of the net accretion rate, 
though it needs to be verified. We thus introduce the idea of two MRI on-set layers in order to discuss the state transitions in more detail. 
Begelman et al. (2015) considered that the MRI on-set condition can also be established in hot coronal layer across the MRI dead zone. 
 In  the dead zone, the gas is heated by the toroidal magnetic flux rising from below and the density decreases drastically. 
 The low density makes the cooling inefficient and electrons and protons are thermally decoupled. 
 Then, the temperature increases very rapidly and the right hand side of Eq. (\ref{eq:10}) overcomes the left-hand side and the MRI can operate again.

The ratio of accretion rates between upper and lower layers is written as
\begin{equation}
\frac{\dot{m}_2}{\dot{m}_1}=\frac{L_2}{L_1}\sim \frac{z_1R}{z_2^2}
\sim 0.7 y^{0.3}\alpha^{0.9}\dot{m}_1^{-0.7},
\label{eq:24}
\end{equation}
(Begelman et al. 2015) where the subscript character 1 and 2 denote the physical quantities of lower and upper layers.  
$y$ is the Compton $y$-parameter and $\dot{m}=\dot{M}/\dot{M}_{\rm{Edd}}$ is the dimensionless accretion rate, $\dot{M}_{\rm{Edd}}=L_{\rm{Edd}}/c^2=4\pi GM_X/\kappa c$ where $L_{\rm{Edd}}$ is the Eddington luminosity.
Moreover, $z_1$ is the scale height of the lower MRI on-set layer and $z_2$ is the height where the on-set condition establishes again across the MRI dead zone. 
Eq. (\ref{eq:24}) is based on the case of the flow near the BH.
We confirmed that the accretion rate of the upper layer in the torus-like flow $\dot{M}_2(R\sim R_c)$ is negligible.
Thus, the net accretion rate of the torus-like flow can be simply written by Eq. (\ref{eq:14}).
Then, assuming that $\dot{M}_{\rm{net}}$ is distributed between the two layers near the BH, we use the ratio $\frac{\dot{m}_2}{\dot{m}_1}$ in Eq. (\ref{eq:24}) to discuss the spectral states.

Begelman et al. (2015) connected the ratio with the spectral states; HS, intermediate HS, intermediate LH and LH states.
However, BH HMXBs traverse a narrower range of the hardness-intensity diagram compared with LMXBs which show the state transitions continuously thorough the two intermediate states (Done et al. 2007; Nowak et al. 2012; Ruhlen et al. 2011).
Cyg X-1 shows the LH-HS (HS-LH) transition with changing hardness and intensity at the same time, and does not show any hysteresis phenomenon.
Thus, we roughly treat the ratio as an indicator so that $\frac{\dot{m}_2}{\dot{m}_1}<1$ for the HS state and $\frac{\dot{m}_2}{\dot{m}_1}>1$ for the LH state. 
We assume that the Compton $y$-parameter $y\sim L_{\rm{tail}}/ L_{\rm{disk}} \sim \dot{m}_2/ \dot{m}_1$ for the HS state where $L_{\rm{tail}}$ and $L_{\rm{disk}}$ are the luminosities of the Compton tail and black body disk components in the X-ray spectrum, while observationally suggested value is $y\sim O(1)$ for the LH state (Begelman et al. 2015).
If we assume $\alpha=0.15$, $\dot{m}_1+\dot{m}_2 = \mbox{constant}$ and the Compton $y$ parameter, we can obtain the accretion rates of each layer.

For Cyg X-1 in the HS state, the accretion rates are $\dot{M}_1\sim3\times 10^{17}$ g s$^{-1}$ and  $\dot{M}_2\sim 1\times 10^{17}$ g s$^{-1}$
 in order to be consistent with the expected mass accretion rate of the HD calculation, i.e. $\dot{M}_{\rm{expected}} = \dot{M}_1 + \dot{M}_2 = 4\times 10^{17}$ g s$^{-1}$ (Table \ref{tab_v4:1}).
Cyg X-1 does not exhibit a purely disk-dominated spectrum without a hard tail.
Although the origin is unknown, the relatively high accretion rate of the coronal flow $\dot{M}_2$ seems to be consistent with the observation.
In the LH state, we obtain $\dot{M}_1\sim3\times 10^{16}$ g s$^{-1}$ and $\dot{M}_2\sim7\times 10^{16}$ g s$^{-1}$.
The disk component is not always observed in the LH state.
Therefore, there may be some other physical mechanisms to explain the observation such as evaporation model (Meyer \& Meyer-Hosmeister 1994; Spruit \& Deufel 2002; Mayer \& Pringle 2007).
However, a detailed discussion about these processes is beyond the scope of this paper.
We note that although the net accretion rate only varies by a factor of 4, $\dot{M}_1$ changes by an order of magnitude and the magnitude correlation between the two layers is reversed because $\dot{M}_2 \propto \dot{M}_1^{0.3}$ and $\dot{M}_{\rm{net}} = \dot{M}_1+\dot{M}_2$. 
Moreover, the luminosity $\sim 10^{37}$ erg s$^{-1}$ is an intermediate value compared with the luminosities of LMXBs in HS state and in LH state.
Thus, we believe that Cyg X-1 stays near the boundary of two states and the reversal could drastically change the X-ray spectrum.

For LMC X-1, we find that $\dot{M}_2 $ is much lower than $\dot{M}_1$ because of the high luminosity $L_X\sim 10^{38}$ erg s$^{-1}$. 
We find again that the net accretion rate must drop by more than an order of magnitude to cause the state transition if parameter $\alpha =0.15$ is assumed.
Thus, it is inferred that Cyg X-1 can undergo a state transition, while it is difficult to trigger the transition in LMC X-1.

  \section{Conclusion}\label{sec:4}
 
 We summarize the results of the present work: 
 \begin{itemize}
\item
Because the BH captures the line-driven wind, it is not clear that the accretion flow physics for disk-fed objects can apply to wind-fed objects in the same way.
 We investigated wind-fed and viscous processes semi-analytically. 
 The captured mass rate becomes higher than the accretion rate expected by the observed luminosity.
\item
 Since the density of captured wind  $\rho\sim 10^{-13}-10^{-14}$ g cm$^{-3}$ is not high enough to be optically thick, the radiation does not become a blackbody.
 Therefore, the wind gas does not directly connect with the multi-color disk.
 The radius of the disk in Cyg X-1 is around an order of magnitude smaller than LMC X-1's which is comparable to LMXBs.
\item
We suggest that the net accretion rate is determined by the MRI on-set condition.
Moreover, the critical radius is independent of the wind velocity.
Instead the velocity makes the difference of features between the two states as the result of the transition.
From these ideas, we conclude that the state transitions are mainly caused by the fluctuation of the critical magnetic field strength $\delta B_c\lesssim B_c$.
Our state transition model can naturally explain the stable feature of BH HMXBs because of the relatively weak dependence of the luminosity on the wind velocity.
\item
Furthermore, our model can also explain the different features between Cyg X-1 and LMC X-1.
 The former undergoes transitions because the radius of the He$\rm{{I\hspace{-.1em}I}}$ ionization front can change drastically and it has a relatively low luminosity.
 The latter does not show transitions because it has an unchanged He$\rm{{I\hspace{-.1em}I}}$ ionization front and a high luminosity. 
 We argue that the physical processes in the outer region of the accretion flow, or the black-hole binary torus, plays an important role in BH HMXBs.

\end{itemize}

\acknowledgments
\section*{Acknowledgements}
The authors would like to thank the referee for his/her valuable comments and Magnus Axelsson for his reading the manuscript and helpful advices.
S.Y is supported by the JSPS Grant-in-Aid for Scientific Research 15H05438, 15H00785 and 16H03954, and KM is by 15K05024.


\end{document}